\documentclass[12pt]{iopart}
\usepackage{subfig}

\usepackage{color}
\usepackage{graphicx}
\usepackage{amssymb}

\def\be{\begin{equation}}
\def\ee{\end{equation}}
\def\ba{\begin{array}}
\def\ea{\end{array}}
\def\beq{\begin{eqnarray}}
\def\eeq{\end{eqnarray}}

\begin{document}
\title{Dissipative dynamics of an open quantum battery}
\author{M Carrega$^{1}$, A Crescente$^{2,3}$, D Ferraro$^{2,3}$, M Sassetti$^{2,3}$ }

\address{$^{1}$NEST, Istituto Nanoscienze-CNR and Scuola Normale Superiore, Piazza S. Silvestro 12, I-56127 Pisa, Italy}

\address{$^{2}$Dipartimento di Fisica, Universit\`a di Genova, Via Dodecaneso 33, 16146, Genova, Italy}
\address{$^{3}$SPIN-CNR, Via Dodecaneso 33, 16146 Genova, Italy}
\ead{matteo.carrega@nano.cnr.it}
\begin{abstract}
Coupling with an external environment inevitably affects the dynamics of
a quantum system. Here, we consider how charging performances of a
quantum battery, modelled as a two level system, are influenced by the presence of an Ohmic thermal reservoir. The latter is coupled to both longitudinal and transverse spin
components of the quantum battery including decoherence and pure dephasing mechanisms.
Charging and discharging dynamics of the quantum battery, subjected to a static driving, are obtained exploiting a proper mapping into the so-called spin-boson model. Analytic expressions for the time evolution of the energy stored in the weak coupling regime are presented relying on a systematic weak damping expansion. Here, decoherence and pure dephasing dissipative coupling are discussed in details. We argue that the former results in better charging performances, showing also interesting features reminiscent of the Lamb shift level splitting renormalization induced by the presence of the reservoir. Charging stability is also addressed, by monitoring the energy behaviour after the charging protocol has been switched off.
This study presents a general framework to investigate relaxation effects, able to include also non Markovian effects, and it reveals the
importance of controlling and, possibly, engineering system-bath coupling in the realization of quantum batteries.
\end{abstract}

\section{Introduction}
\label{sec:introduction}
Quantum properties can play a predominant role in determining the behaviour of micro- and nano-devices. Recently, both theoretical and experimental works considered thermodynamic aspects of small quantum systems, in the new research field called ``quantum thermodynamics''~\cite{Esposito09, Levy12, Pekola15, Vinjanampathy16, Benenti17, Depasquale18, Bera19, Carrega19}.
In this context one of the major issues, also triggered by potential technological applications, is the possibility to efficiently store energy in small systems, exploiting quantum features, and using it on-demand providing power supply. This naturally leads to the idea of \emph{quantum batteries} (QBs)~\cite{Campaioli18}, devices where the performances in terms of energy transfer and charging power, namely the energy stored or released in a given time interval, can be improved by exploiting quantum resources such as entanglement~\cite{Alicki13, Hovhannisyan13, Binder15, Campaioli17}. On the one hand the attention focused on the characterization of possible quantum advantage of single~\cite{Binder15, Julia-Farre18, Zhang19, Chen19, Crescente20} and many-body QBs~\cite{Le18, Rossini19, Rosa19} over their classical counterparts~\cite{Andolina19b}.
On the other hand, theoretical frameworks aiming at actual experimental implementations in several setups such as circuit-QED, already used for quantum computing purposes~\cite{Devoret13}, are under investigation~\cite{Ferraro18}.
Here, a paradigmatic playground is a quantum two-level systems (TLS)~\cite{Wiel02, Koch07}, physically realized by means of superconducting qubits~\cite{Devoret13} or quantum dots~\cite{Singha11} in semiconducting nanostructures.
This, indeed, represents the elementary building block (cell) for realizing QBs.
Charging of a TLS, namely the controlled transition between the ground and the excited state, can be induced by means of an external classical drive~\cite{Zhang19, Chen19, Crescente20}, by properly controlling the exchange interaction between different cells~\cite{Campaioli18, Le18, Rossini19}, or through cell-cell coupling mediated by interaction with an external cavity radiation~\cite{Ferraro18, Andolina19, Ferraro19, Andolina18}.  

Most of the literature on QBs focused on the dynamics of closed systems where the energy is coherently transferred from a charger to the battery~\cite{Campaioli18, Binder15, Campaioli17, Andolina18}, leaving only marginal discussions on possible effects due to the presence of external environment.
However, as it is well known, the unavoidable coupling with an environment  is responsible for dissipation, leading to relaxation and dephasing of each two-level system~\cite{Haroche_Book, uli_book} and therefore should be properly addressed also in this context.
First investigations on ``open'' quantum batteries have been based on the study of the time evolution of the reduced density matrix of the TLS following a conventional Lindblad approach of Markovian master equations~\cite{Ou17, Farina19}. Within this framework, and under Markov approximation, general constraint on the possible stored energy and averaged charging power associated to the QB have been recently introduced~\cite{Zakavati20, Kamin20}. Moreover, some protocols, based on multiple projective or weak measurements, able to mitigate these detrimental effects on the efficiency of the QB, have been proposed~\cite{Barra19, Santos19, Gherardini20, Quach20}. However, a microscopic description of the physical processes involved in the energy dissipation and their impact on the performances of the QB is still lacking.

The aim of this work is to investigate how environment-induced dissipation can affect charging (and discharging) dynamics of a quantum battery. To avoid complications due to collective behaviours and interactions between the single entities of a large quantum battery, that can lead to competing effects~\cite{Campaioli18, Julia-Farre18}, we will assume the cells to be independent quantities.
We therefore inspect the dynamics of a single cell quantum battery (hereafter indicated as QB) when a static external classical drive is acting as a charger and in presence of a coupling to an external reservoir responsible for dissipation. We underline that several system-reservoir couplings have been investigated and engineered, depending on the actual implementation of the quantum device under study~\cite{Devoret13, uli_book, schnirman_2002, vion_2002}. Among all, we will concentrate on the case of linear dissipative couplings described by an Ohmic spectrum, which well describes the low energy noise source of many solid-state devices~\cite{uli_book, schnirman_2002}.

Dissipative dynamics of open QB is systematically investigated by mapping the problem in the so-called spin-boson (SB) model~\cite{uli_book, schnirman_2002, brandes_2002, caldeira_1983,  leggett_1987}.
As one would expect, too strong dissipation inevitably lead to fast incoherent relaxation dynamics, undermining the potential use of TLS as quantum batteries. Therefore, in the following, specific results will be discussed in the case of weak dissipation strength (weak damping regime), where exact analytic results can be derived.
Focussing on the average energy variation, we investigate both charging and discharging dynamics, underlining the different behaviour between decoherence and pure dephasing linear dissipative couplings. By identifing the energy flows, both in the charging and discharging process, we will show that the former possible coupling results in better QB performances at fixed dissipation strength.\\ 

The paper is organized as follows. In Section~\ref{sec:model} we present the model of a single cell QB coupled to a reservoir (thermal bath). Different QB-bath couplings, leading to decoherence or pure dephasing mechanisms, are taken into account, by means of proper mapping to the SB model.
Section~\ref{sec:master} focuses on the dissipative dynamics, recalling generalized master equations governing the time evolution of the TLS in the presence of Ohmic dissipation. Here, analytic closed expressions for the average energy associated to the QB in the weak damping regime are presented.
Finally, Section~\ref{sec:results} contains our main results on charging performances in presence of dissipation and Section~\ref{sec:conclusions} summarizes our main conclusions.
Technical details, together with a discussion of the effect due to stronger dissipation, are reported in three Appendices.

\section{Open quantum battery}
\label{sec:model}
We consider a single cell QB described by the Hamiltonian (hereafter we set $\hbar=1$)
\be
\label{eq:hqb}
H_{{\rm QB}}=\frac{\Delta}{2}\sigma_z~,
\ee
where $\Delta$ represents the level spacing between the ground and the excited state, which can be seen as the empty and the full cell configuration respectively.
At time $t=0^+$ a coupling with the $\sigma_x$ component of the QB with an external classical field $A$ (the charger) is switched on~\cite{Binder15, Campaioli17, Ferraro18} 
\be
\label{eq:hcharger}
H_{{\rm C}}=\frac{A}{2}\sigma_x~.
\ee
In the above equations $\sigma_k$ ($k=x,y,z$) indicates the usual $k$-th Pauli matrix.
The QB is also coupled to a reservoir (thermal bath), responsible for dissipation, modelled as an ensemble of harmonic oscillators of frequency $\omega_j$~\cite{uli_book, caldeira_1983, leggett_1987, ingold_2002}. In terms of bosonic creation (annihilation) operators $a^{\dagger}_j (a_j)$ it reads 
\be
\label{eq:hreservoir}
H_{{\rm R}}=\sum_{j}\omega_{j} a^{\dagger}_ja_j ~.
\ee
We consider a linear coupling with the reservoir along both the longitudinal ($z$) and transverse ($x$) directions
\be
\label{eq:hint}
H_{{\rm I}}=\frac{1}{2}\left[\sigma_x\cos(\theta/2) + \sigma_z\sin(\theta/2) \right]\cdot\sum_{j}\lambda_j\left(a^{\dagger}_j+a_j\right),
\ee
capturing both decoherence ($\theta=0$) and pure dephasing ($\theta=\pi$) processes~\cite{schnirman_2002, gramich_2011, gramich_2014}. Notice that different couplings can correspond to different and independent noise sources, for instance they can be linked to charge and flux noise in a superconducting Josephson realization of a TLS~\cite{schnirman_2002, vion_2002, makhlin_2004}.
The spectral properties of the reservoir are characterized  by the density function~\cite{uli_book}
\be
\label{spectraldefinition}
 J(\omega) = \sum_j \lambda_j^2 \delta (\omega -\omega_j),
 \ee
which  in the continuum limit and in the relevant case of Ohmic dissipation we are interested in, has the form~\cite{uli_book,  leggett_1987,ingold_2002} 
\be
  J(\omega) =2 \alpha \omega\, \e^{-\omega/\omega_{ c}}.
\label{ohmic} 
\ee
Here,  $\alpha$ is a dimensionless parameter which quantifies dissipation strength and $\omega_c$ the high frequency cut-off of the bath~\cite{uli_book, sassetti_1990, grifoni_1996}.
 The total Hamiltonian governing the dynamics at $t\geq 0$ is thus given by
\be
\label{eq:htot}
H= H_{{\rm QB}} + H_{{\rm C}} + H_{{\rm RI}}~,
\ee
where we have defined the total contribution of the reservoir $H_{{\rm RI}}=H_{{\rm R}}+H_{{\rm I}}$.
At time $t=0$, we assume factorized initial  conditions, with a total density matrix given by $\rho_{{\rm tot}}(0)=\rho(0)\times \rho_{{\rm R}}(0)$. Here, the reservoir is at thermal equilibrium with density $\rho_{{\rm R}}(0)=e^{-\beta H_{{\rm R}}}/{\rm Tr}[e^{-\beta H_{{\rm R}}}]$, where $\beta=1/(k_{{\rm B}}T)$ is the bath inverse temperature.
On the other hand, the QB is described by the reduced density matrix 
\beq
\label{eq:rhoqb}
\rho(0)=\left(\begin{array}{cc}
{p}_R & {a}-i{b} \\
{a} +i{b} & {p}_L
\end{array}\right),
\eeq
where we have included the possibility to have initial coherences  (${a}$ and  ${b}$). Normalization of $\rho(0)$ imposes ${p}_R +{p}_L=1$. In addition, with  ${a}$ and  ${b}$ real coefficients, the condition ${\rm Tr}\rho^2\le 1$ leads to the constraint ${a}^2+{b}^2\le {p}_R {p}_L$.
Time evolution of the spin components $\langle\sigma_k(t)\rangle$ ($k=x,y,z$) are written as averages over the time dependent  total density matrix,  $\rho_{{\rm tot}}(t)$ driven by the total Hamiltonian $H$
\be
\label{eq:tracetau}
\langle \sigma_k(t)\rangle= {\rm Tr}[\rho_{{\rm tot}}(t)\sigma_k]~.
\ee
These averages can be represented  in terms of the time dependent reduced density matrix $\rho(t)$ as 
\be
\langle\sigma_k(t)\rangle={\rm Tr}[\rho(t)\sigma_k]~,
\ee
where $\rho(t)={\rm Tr}_{{\rm R}}[\rho_{{\rm tot}}(t)]$ with ${\rm Tr}_{{\rm R}}$ the trace over the bath degrees of freedom.

 Energy exchanges between the different subparts, and thus charging dynamics, at time $t$ can be then determined by studying the energy variations  
\be
\label{eq:energyvariation}
\langle E_{s} (t)\rangle = \langle H_s(t)\rangle -\langle H_s(0)\rangle~,
\ee
where $s={\rm QB}, {\rm C}, {\rm RI}$. Recall that for $t>0$, because of the static driving, we have $\dot{H} =0$, with an energy balance of the form
\be
\langle E_{{\rm QB}}(t)\rangle + \langle E_{{\rm C}}(t)\rangle +\langle E_{{\rm RI}}(t)\rangle=0
\label{eq:energybalance}
\ee
that must hold for any driving amplitude and dissipation coupling strength.\\

Before discussing the dissipative dynamics, it is instructive to recall that in absence of dissipation ($\alpha=0$), all energy supplied by $H_{{\rm C}}$ is transfered to the QB, whose maximum energy that can be stored is given by the energy level spacing $\Delta$.
A static driving protocol then charges the QB according to~\cite{Binder15, Crescente20}
\be
\label{eq:closed}
\langle E_{{\rm QB}}(t)\rangle =\frac{\Delta}{2}\frac{A^2}{\Omega^2}\big[1-\cos(\Omega t)\big]~,
\ee
where $\Omega$ is the Rabi frequency
\be
\label{eq:rabi}
\Omega=\sqrt{\Delta^2+A^2}~.
\ee

Here, the unitary evolution of the closed system implies a periodic behaviour, as evident from Equation~(\ref{eq:closed}),
with the QB completely discharged (again empty and in the $|g\rangle$ state) for even multiplies of $\Omega t= 2\pi n$ ($n>0$ integer).
Notice that the maximum amount of the energy stored, is reached for very large amplitude $A\gg \Delta$. 
\subsection{Mapping to the spin boson model}
\label{sec:mapping}

In order to study the time evolution of the different subparts we start by considering a unitary rotation in the spin space 
\be
{\cal R}=e^{-i\phi\sigma_y}
\ee
with the angle  $\phi$ chosen in such a way to project the interaction part $H_{\rm I}$ only along the $z$ axis. It is easy to show (see~\ref{app:a} for details) that with $\phi=(\theta+\pi)/4$ one has
\beq
{\widetilde H}_{\rm I}&\equiv&\mathcal{R}H_{\rm I}\mathcal{R}^{-1}=-\frac{1}{2}\sigma_z\cdot\sum_{j}\lambda_j\left(a^{\dagger}_j+a_j\right),
\label{interaction}\\
{\widetilde H}_{\rm QB}&\equiv&\mathcal{R}H_{\rm QB}\mathcal{R}^{-1}=-\frac{\Delta}{2}\left[\sigma_z\sin(\theta/2)-\sigma_x\cos(\theta/2)  \right]\label{QB}\\
\label{C}
{\widetilde H}_{\rm C}&\equiv&\mathcal{R}H_{\rm C}\mathcal{R}^{-1}=-\frac{A}{2}\left[\sigma_z\cos(\theta/2)+\sigma_x\sin(\theta/2)  \right].
\eeq
In this way, the  total Hamiltonian 
\be {\widetilde H}={\widetilde H}_{\rm QB}+{\widetilde H}_{\rm C}+H_R+{\widetilde H}_{\rm I}
\label{widetildeH}
\ee
can be then recast in the standard form of a SB model~\cite{uli_book, henriet_2014}, which represents a prototypical quantum dissipative two state system that has been studied with several numerical~\cite{bulla_2003, orth_2010, thorwart_2015, palm_2018, stockburger_2004, stockburger_2016} and analytical approaches both in the weak and strong coupling regimes~\cite{uli_book, sassetti_1990, grifoni_1996, grifoni_1999, grifoni_1997, carrega_2015, carrega_2016}.
We have indeed
\be
{\widetilde H}\equiv H^{{(\rm SB)}}=H^{(SB)}_0+ H_{{\rm R}} + H^{(SB)}_{{\rm I}}
\label{eq:h-sb}
\ee
where
\be
\label{eq:sb_0}
H^{(SB)}_0= -\frac{\Delta_0}{2}\sigma_x-\frac{\epsilon_0}{2}\sigma_z,
\ee
is the bare TLS  Hamiltonian, with $\Delta_0$ the tunneling amplitude and $\epsilon_0$ a constant bias, and  $H^{(SB)}_{{\rm I}}= {\widetilde H}_{{\rm I}}$
given in Equation~(\ref{interaction}). By comparing the above $H^{{(\rm SB)}}$ with 
${\widetilde H}$ in Equation~(\ref{widetildeH}) we obtain the following identifications
\beq
\label{eq:delta0}
\Delta_0&=&\left[A\sin(\theta/2)-\Delta\cos(\theta/2)\right]\,\\
\epsilon_0&=&\left[A\cos(\theta/2)+\Delta\sin(\theta/2)\right].
\label{eq:epsilon0}
\eeq
To complete the mapping we still need to identify the reduced SB initial  density matrix, as
\beq
\rho_{\rm SB}(0)=\left(\begin{array}{cc}
\bar{p}_R & \bar{a}-i \bar{b} \\
\bar{a}+i \bar{b} & \bar{p}_L
\label{initial}
\end{array}\right)\,,
\eeq
with  the rotated reduced initial matrix $\widetilde\rho(0)\equiv \mathcal{R}\rho(0)\mathcal{R}^{-1}$ of the QB. We therefore get (see~\ref{app:a}) 
\beq
\label{pr}
\bar{p}_R&=&-p_R\sin(\theta/2)+\frac{1+\sin(\theta/2)}{2}-a\cos(\theta/2),\\
\bar{p}_L &=&-p_L\sin(\theta/2)+\frac{1+\sin(\theta/2)}{2}+a\cos(\theta/2),\\
\bar{a}&=&-a\sin(\theta/2)+\frac{p_R-p_L}{2}\cos(\theta/2),\\
\label{b}\bar{b}&=&b.
\eeq\\
From now on, the average  energy of the QB,
$\langle H_{\rm QB}(t)\rangle$ and that of the charger $\langle H_{\rm C}(t)\rangle$
will be directly obtained using the expressions (\ref{QB}) and (\ref{C}) as
\beq
\langle H_{\rm QB}(t)\rangle&=&-\frac{\Delta}{2}\left[\langle\sigma_z(t)\rangle\sin(\theta/2)-\langle\sigma_x(t)\rangle\cos(\theta/2)\right]\label{HQB}\\
\langle H_{\rm C}(t)\rangle&=&-\frac{A}{2}\left[\langle\sigma_z(t)\rangle\cos(\theta/2)+\langle\sigma_x(t)\rangle\sin(\theta/2)\right]\label{HC}
\eeq
where the time evolution of the spin components will be considered in the SB model as $\langle\sigma_k(t)\rangle={\rm Tr}[\rho_{{\rm SB}}(t)\sigma_k]$ with the proper identifications just discussed.

\section{Dissipative dynamics}
\label{sec:master}

\subsection{Generalized master equations}
\label{GME}
As mentioned above, it is convenient  to solve the dynamics of the QB by using the mapping to the SB model 
given in Equation~(\ref{eq:h-sb})  with  initial  density matrix  in Equation~(\ref{initial}). Indeed, for this model, the time dependent evolution  of the spin components $\sigma_{z}(t)$ and $\sigma_{x}(t)$, were studied with several numerical and analytical methods~\cite{uli_book, orth_2010, stockburger_2004, carrega_2015, carrega_2016, orth_2013}.
In particular, in the framework of the path integral approach~\cite{uli_book, brandes_2002, ingold_2002, grifoni_1999, carrega_2015} it was 
possible to represent $\langle \sigma_{z}(t)\rangle$ in the form of an exact generalized master equations (GME), and to connect $\langle \sigma_{x}(t)\rangle$ with $\langle \sigma_{z}(t)\rangle$ by an exact integral relation~\cite{uli_book, grifoni_1996, grifoni_1997}. Usually in this approach the reduced initial density matrix is choosen to be diagonal, with $\bar{a}=\bar{b}=0$. 
Here, following the procedure outlined in \cite{grifoni_1999}, we will relax this assumption  by considering also the presence of non-diagonal terms, which are necessary for the mapping to the QB. Below we will briefly present the main steps, while the details are reported in~\ref{app:kernels}.

Let us start with the $z$ spin component which fulfills an exact integro-differential equation
\be
\hspace{-1cm}\frac{d\langle \sigma_{z}(t)\rangle }{dt}= \int_0^t dt' [K^{(-)}_{1,z} (t-t') - K^{(+)}_{1,z}(t-t')\langle \sigma_{z}(t')\rangle ] +2\bar{a} K^{(-)}_{2,z}(t)-2\bar{b} K^{(+)}_{2,z}(t),
\label{eq:gme_sigmaz}
\ee
with initial condition $\langle \sigma_{z}(0)\rangle= \bar{p}_R-\bar{p}_L$. The kernels $K_{1,z}^{(\pm)}(t-t')$ determine  the spin evolution in the presence of an initial diagonal density matrix  ($\bar a=\bar b=0$), on the other hand,  $K_{2,z}^{(\pm)}(t-t')$ 
are responsible for the  additional contributions due to the initial coherence terms ($\bar a\not= 0$, $\bar b\not= 0$). The upper label $(+)$ and $(-)$ indicate whether the kernel is an even or odd function of the bias $\epsilon_0$. 
All these kernels encorporate dissipative effects and they only depend on the time difference since the SB Hamiltonian is time independent. 

As shown in~\ref{app:kernels} they are expressed in terms of an infinite series  over all possible tunneling processes, governed by $\Delta_0$.
Notice that,  due to the linearity of Equation~(\ref{eq:gme_sigmaz}),  $\langle\sigma_z(t)\rangle$ can be always decomposed as
\be
\label{eq:sz_sum}
\langle\sigma_z(t)\rangle=\langle\sigma_{z,0}(t)\rangle+\langle\sigma_{z,\bar{a}}(t)\rangle+\langle\sigma_{z,\bar{b}}(t)\rangle,
\ee
where $\langle\sigma_{z,0}(t)\rangle$ is the contribution without the initial coherence terms ($\bar{a}\!=\!\bar{b}\!=\!0$),
while $\langle \sigma_{z,\bar{a}}(t)\rangle $ and $\langle \sigma_{z,\bar{b}}(t)\rangle $ 
are the terms due to the presence of the coefficients $\bar{a}$ and $\bar{b}$ respectively. 
As demonstrated in~\ref{app:kernels} (see Equation (\ref{Bgenerallinks1})) in the so called scaling limit, i.e. large cut-off frequency $\omega_c$, these two parts can be directly linked  to $\langle\sigma_{z,0}(t)\rangle$ in the following way
\begin{eqnarray}
\langle\sigma_{z,\bar{a}}(t)\rangle&=& \frac{2\bar{a}}{\Delta_0\sin(\pi\alpha)}\frac{d}{dt}\langle{\sigma}^{(-)}_{z,0}(t)\rangle\nonumber\\
\langle\sigma_{z,\bar{b}}(t)\rangle&=& \frac{2\bar{b}}{\Delta_0\cos(\pi\alpha)(\bar{p}_R-\bar{p}_L)}\frac{d}{dt}\langle{\sigma^{(+)}_{z,0}(t)}\rangle,
\label{generallinks1}
\end{eqnarray}
where again $\pm$ indicates symmetric/antisymmetric term with respect to $\epsilon_0$.
It is important to underline that this  result is valid at any order in the dissipation coupling strength $\alpha$ and in  the tunneling amplitude $\Delta_0$. These  relations are particularly helpfull since they allow to evaluate the full dynamics of  $\langle\sigma_z(t)\rangle$  starting from initial diagonal conditions  ($\bar{a}=\bar{b}=0$), and deriving the full expressions also in the presence of coherent (off-diagonal) components of the initial density matrix. 

Let us comment now on the general structures of $\langle \sigma_{x}(t)\rangle$. As shown in~\cite{grifoni_1999, grifoni_1997} this quantity is directly connected to $\langle \sigma_{z}(t)\rangle$ via an exact integral relation
\be
\hspace{-1cm}\langle \sigma_{x}(t)\rangle = \int_0^t dt' [K^{(+)}_{1,x}(t-t') + K^{(-)}_{1,x}(t-t')\langle \sigma_{z}(t')\rangle ]  +2\bar{a} K^{(+)}_{2,x}(t)+2\bar{b} K^{(-)}_{2,x}(t).
\label{eq:gme_sigmax}
\ee
The Kernels $K_{1/2,x}^{(\pm)}$ are again given in the form of a series  and they are quoted in~\ref{app:kernels} (see Equation (\ref{Bkx})). 
Also this quantity can be decomposed as a sum of a term without  $\bar{a}$ and $\bar{b}$, called  $\langle\sigma_{x,0}(t)\rangle$ and the remaining parts  $\langle \sigma_{x,\bar{a}}(t)\rangle$ and   $\langle \sigma_{x,\bar{b}}(t)\rangle$ as
\be
\label{eq:sx_sum}
\langle\sigma_x(t)\rangle=\langle\sigma_{x,0}(t)\rangle+\langle\sigma_{x,\bar{a}}(t)\rangle+\langle\sigma_{x,\bar{b}}(t)\rangle.
\ee
Similarly to the $z$-component,   for sufficiently large cut-off frequency, the two last terms are linked to  $\langle\sigma_{x,0}(t)\rangle$ as
\begin{eqnarray}
\langle\sigma_{x,\bar{a}}(t)\rangle&=& \frac{2\bar{a}}{\Delta_0\sin(\pi\alpha)}\frac{d}{dt}\langle{\sigma^{(+)}_{x,0}(t)}\rangle\nonumber\\
\langle\sigma_{x,\bar{b}}(t)\rangle&=& \frac{2\bar{b}}{\Delta_0\cos(\pi\alpha)(\bar{p}_R-\bar{p}_L)}\frac{d}{dt}\langle{\sigma^{(-)}_{x,0}(t)}\rangle.
\label{generallinks2}
\end{eqnarray}
Few remarks on the possible approaches to solve the above exact expressions are now in order. In general  the complete resummation of the series expansions which determine the kernels $K_{1/2,z}^{(\pm)}$ and $K_{1/2,x}^{(\pm)}$  cannot be done in closed form.
 Therefore one needs to resort to suitable analytical approximations or numerical computations.  A well-known example is the so-called non-interacting blip approximation (NIBA)~\cite{uli_book, brandes_2002,  orth_2013,  hartmann_2000}. As shown in the final part of~\ref{app:kernels} this scheme approximates the kernels $K_{1/2,z/x}^{(\pm)}$  at their lowest order in $\Delta_0$ truncating then their series expansions. On the other hands,  it  retains the dissipation strength $\alpha$ at any order.  It has been shown that NIBA is a good approximation for sufficiently high temperatures $\beta\sqrt{\Delta_0^2 + \epsilon_0^2}<1 $, while strong deviations  occur at low temperatures especially  in the long time regime and in the presence of a finite bias~\cite{uli_book}.
Another, complementary, approximation  is the so called systematic weak damping expansion, valid for $\alpha\ll1$. This method evaluates the kernels at lowest order in $\alpha$ with an exact  resummation of the series in $\Delta_0$. This is a very powerful method to treat dissipative dynamics in the weak coupling regime at low temperature $\beta\sqrt{\Delta_0^2+\epsilon_0^2}>1$, where quantum  coherence and non markovianity effects may play an important role. 
\subsection{Weak damping dynamics}
\label{app:weak_sb}

Here we specify the spin dynamics  in the weak damping regime in the SB model. In the limit $\alpha \ll1$ and low temperature regime $\beta\sqrt{\Delta_0^2+\epsilon_0^2}>1$ it is possible to obtain closed analytical expressions for the time evolution of spin operators relying on a systematic weak damping espansion, valid at any order in the tunneling amplitude $\Delta_0$. Resummation of the infinite series expressions in the systematic weak damping expansion have been usually considered starting from diagonal initial conditions, neglecting coherence terms~\cite{uli_book}. However, by exploiting the general links derived in~\ref{app:kernels}, from the knowledge of $\langle \sigma_{z,0}(t)\rangle $ and $\langle \sigma_{x,0}(t)\rangle $ it is easy to derive also the expressions related to finite off diagonal contributions $\bar{a}$ and $\bar{b}$.
We briefly remind the main steps behind this systematic expansion, for the $\langle \sigma_{z,0}(t)\rangle$ contribution. The starting point is the formal series expression (reported in ~\ref{app:kernels}) of the kernels entering the generalized master equation~(\ref{eq:gme_sigmaz}) $K_{1,z}^{(\pm)}(t-t')$.
The sum in Equation~(\ref{Bkappa}) can be resummed at all order in $\Delta_0$, considering the lowest order expansion in $\alpha$ of the exponential factors present in~(\ref{eq:bg1}) which enters in Equation~(\ref{eq:bf}). Notice that in this approach, the kernels $K_{1,z}^{(\pm)}$ can be viewed as self-energy correction for the operator $\langle \sigma_{z,0}(t)\rangle$. Upon the expansion at lowest order in $\alpha$ the series expression for the kernels can be summed and the resulting expression can be plugged into the generalized master equation which now admits a simple closed solution.
Following a similar procedure it is possible to obtain a systematic weak damping expression also for the $x$- component~\cite{uli_book, grifoni_1997}.

The averaged spin component $\langle \sigma_{k,0}(t)\rangle$ with $k=z,x$ can be written in analytic form as
\be
\hspace{-1cm}\langle\sigma_{k,0}(t)\rangle=  N^{(1)}_{k,0} e^{-\Gamma^{(r)} t} + \left[N^{(2)}_{k,0} \cos(\Omega_{{\rm SB}} t) + N^{(3)}_{k,0} \sin (\Omega_{{\rm SB}} t)\right] e^{-\Gamma t}+ \langle\sigma_k(\infty)\rangle.
\label{eq:sigma0_weak}
\ee
Both quantities posses an oscillatory behaviour, with characteristic frequency
\be
\Omega_{{\rm SB}}=\sqrt{ \Delta_{0,{\rm eff}}^2+\epsilon_0^2},
\ee
with
\be
\Delta_{0,{\rm eff}}=\Delta_0 \left[\frac{\Delta_0}{\omega_c}\right]^{\alpha/(1-\alpha)}\cdot\left[\Gamma(1-2\alpha)\cos(\pi\alpha)\right]^{1/[2(1-\alpha)]}~.
\ee
The level splitting gets  renormalized with respect to the bare case ($(\Delta_0^2+\epsilon_0^2)^{1/2}$), with always $\Omega_{{\rm SB}}<\sqrt{\Delta_0^2+\epsilon_0^2}$.
The oscillatory behaviour present in Equation~(\ref{eq:sigma0_weak}) is modulated by exponential decay dictated by the incoherent relaxation $\Gamma^{(r)}$ and dephasing $\Gamma$ rates, given by
\begin{eqnarray}
\label{eq:rates}
\Gamma^{(r)} &=& \frac{\pi \alpha \Delta_{0,{\rm eff}}^2}{\Omega_{{\rm SB}}}\coth[\beta \Omega_{{\rm SB}}/2]\nonumber\\
\Gamma&=&\frac{\Gamma^{(r)}}{2} + 2\pi \alpha \frac{\epsilon_0^2}{\beta \Omega_{{\rm SB}}^2}.
\end{eqnarray}
The amplitudes entering the above expressions are evaluated up to linear order in $\alpha$, (apart the renormalized frequency $\Delta_{0,{\rm eff}}$) and read

\begin{eqnarray}
N^{(1)}_{z,0}&=& \frac{(\bar{p}_R-\bar{p}_L)\epsilon_0^2}{\Omega_{{\rm SB}}^2} - \langle\sigma_z(\infty)\rangle \nonumber\\
N^{(2)}_{z,0} &= &\frac{ (\bar{p}_R-\bar{p}_L)\Delta_{0,{\rm eff}}^2}{\Omega_{{\rm SB}}^2}\nonumber\\
N^{(3)}_{z,0}&= & \frac{\Gamma^{(r)} N^{(1)}_{z,0} +\Gamma N^{(2)}_{z,0}}{\Omega_{{\rm SB}}}\nonumber\\
N^{(1)}_{x,0} &=& \frac{\epsilon_0\Delta_{0,{\rm eff}}}{\Omega_{{\rm SB}}^2} (\bar{p}_R-\bar{p}_L) - \langle\sigma_x(\infty)\rangle\nonumber\\
N^{(2)}_{x,0} &=&- \frac{\epsilon_0\Delta_{0,{{\rm eff}}}}{\Omega_{{\rm SB}}^2} (\bar{p}_R-\bar{p}_L)\nonumber\\ 
N^{(3)}_{x,0} &=&\frac{1}{\Omega_{{\rm SB}}}\left[\Gamma^{(r)} N^{(1)}_{x,0} +\Gamma N^{(2)}_{x,0}+\pi\alpha\Delta_{0,{\rm eff}}\right].
\label{expression}
\end{eqnarray}
In the above equations we have introduced  the steady ($t\to\infty$) values  
\begin{eqnarray}
\label{eq:steady}
\langle\sigma_z(\infty)\rangle&=& \frac{\epsilon_0}{\Omega_{{\rm SB}}}\tanh [\beta\Omega_{{\rm SB}}/2]\nonumber\\
\langle\sigma_x(\infty)\rangle&=&\frac{\Delta_{0,{\rm eff}}}{\Omega_{{\rm SB}}}\tanh[\beta\Omega_{{\rm SB}}/2].
\end{eqnarray}
Exploiting Equation~(\ref{Bgenerallinks1})-(\ref{Bgenerallinks2}), with the substitution $\Delta_0\to \Delta_{{\rm eff}}$ to take consistently into account the dissipation induced renormalization, we obtain 
\beq
\hspace{-2cm}\langle\sigma_{z,\bar{a}}(t)\rangle&=& \frac{2\bar{a} \Delta_{0,{\rm eff}}\epsilon_0}{\Omega_{{\rm SB}}^2}\Big[e^{-\Gamma^{(r)} t} -e^{-\Gamma t} \cos(\Omega_{{\rm SB}} t)
+ \frac{\Gamma}{\Omega_{{\rm SB}}}e^{-\Gamma t}\sin(\Omega_{{\rm SB}} t)\Big]\label{zabar}
\\
\hspace{-2cm}\langle\sigma_{z,\bar{b}}(t)\rangle&=& \frac{-2\bar{b}}{\Delta_{0,{\rm eff}}\Omega_{{\rm SB}}^2}\Big\{
\epsilon_0^2\Gamma^{(r)}[e^{-\Gamma^{(r)} t} - e^{-\Gamma t}\cos(\Omega_{{\rm SB}} t)] +\Delta_{0,{\rm eff}}^2\Omega_{{\rm SB}} e^{-\Gamma t}\sin(\Omega_{{\rm SB}} t)\Big\} \label{zbbar} \\
\hspace{-2cm}\langle \sigma_{x,\bar{a}}(t)\rangle &=&\frac{2 \bar{a}}{\Omega_{{\rm SB}}^2} \Big[\Delta_{0,{\rm eff}}^2 e^{-\Gamma^{(r)} t} 
+ \epsilon_0^2e^{-\Gamma t}\cos(\Omega_{{\rm SB}} t) -\frac{\epsilon_0^2\Gamma}{\Omega_{{\rm SB}}} e^{-\Gamma t}\sin(\Omega_{{\rm SB}} t)\Big]\label{xabar}\\
\hspace{-2cm}\langle \sigma_{x,\bar{b}}(t)\rangle &=& -\frac{2 \bar{b}}{\Omega_{{\rm SB}}^2} \Big\{\epsilon_0\Gamma^{(r)} [e^{-\Gamma^{(r)} t} 
- e^{-\Gamma t}\cos(\Omega_{{\rm SB}} t)] - \epsilon_0\Omega_{{\rm SB}}\sin(\Omega_{{\rm SB}} t)e^{-\Gamma t}\Big\}\label{xbbar}.
\eeq
Finally the full time-evolution of $\langle \sigma_z(t)\rangle $ and $\langle \sigma_x(t)\rangle$ in the weak damping regime is obtained by plugging the above expressions 
(Equations (\ref{eq:sigma0_weak}),(\ref{zabar})-(\ref{xbbar})) into Equations~(\ref{eq:sz_sum}) and (\ref{eq:sx_sum}).

\section{Results and discussions}
\label{sec:results}

\subsection{Charging dynamics in the weak damping regime}
\label{sec:weak}
We now use the previous results in order to describe the dynamics of the QB. We focus on the weak damping limit in the low temperature regime, which seems to be the most promising regime in order to still achieve good performance of the QB, even in presence of a thermal bath. Moreover, we note that, in this regime, possible effects related to quantum coherences and non-Markovian contributions can play relevant role in determining the QB dynamics. We will determine the energy variation associated to the charging process  considering the QB initially prepared in the ground state $|g\rangle$, { i.e.} $p_L=1$ and $p_R=a=b=0$. Similar expressions can be obtained considering other initial conditions as well. Unless otherwise stated,  we will  discuss the two limiting case of decoherence coupling ($\theta=0$ in Equation~(\ref{eq:hint})) and of pure dephasing coupling ($\theta=\pi$ in Equation~(\ref{eq:hint})). To distinguish between the two we introduce an index $_{\theta}$ with $\theta=0$ or $\theta=\pi$ whenever necessary. Using the expressions (\ref{HQB}), (\ref{HC}) with the mapping in (\ref{eq:delta0}), (\ref{eq:epsilon0})   the average energies associated to the QB and to the charger C are directly expressed in terms of the time evolution of the spin components $\langle\sigma_{z}(t)\rangle$ and $\langle\sigma_{x}(t)\rangle$ of the SB model.
As explained above their evolution can be exactly solved for weak damping (see Section~\ref{app:weak_sb}). Hereafter, for sake of clarity, we directly quote the closed expressions for the time evolution in terms of QB variables.

We start considering the case of pure decoherence with $\theta=0$ in Equation~(\ref{eq:hint}), where the thermal bath is coupled only to $\sigma_x$ operator. We have
\beq
\label{eq:eqb_x}
&&\hspace{-1.6cm}\langle E_{{\rm QB}}(t)\rangle = \frac{\Delta}{2} \left\{1-\frac{\Delta_{\rm eff}}{\Omega_{0}}\tanh(\frac{\beta\Omega_{0}}{2})+\frac{\Delta_{\rm eff}}{\Omega_{0}}\bigg(\tanh(\frac{\beta\Omega_{0}}{2})-\frac{\Delta_{\rm eff}}{\Omega_{0}}\bigg)e^{-\Gamma_{0}^{(r)} t}+\right. \nonumber \\
&&\hspace{-1.6cm}+\frac{e^{-\Gamma_{0} t}}{\Omega_0^2}\left.\bigg[-A^2\cos(\Omega_{0}t)
+ \bigg(\Delta_{\rm eff}\Gamma_{0}^{(r)}\tanh(\frac{\beta\Omega_{0}}{2})+\frac{A^2\Gamma_{0}}{\Omega_{0}}\bigg)\sin(\Omega_{0}t)\bigg]
\right\},
\eeq
while the energy variation of the charger $\langle E_{{\rm C}}(t)\rangle $ is
\beq
\label{eq:ec_x}
&&\hspace{-1.6cm}\langle E_{{\rm C}}(t)\rangle =\frac{A^2}{2\Omega_{0}}\left\{-\tanh(\frac{\beta\Omega_{0}}{2})+
\bigg[\tanh(\frac{\beta\Omega_{0}}{2})-\frac{\Delta_{\rm eff}}{\Omega_{0}}\bigg]e^{-\Gamma_{0}^{(r)} t}+\right.\nonumber 
\\
&&\hspace{-1.6cm}+\left.\frac{e^{-\Gamma_{0} t}}{\Omega_{0}}\bigg[\Delta_{\rm eff}\cos(\Omega_{0}t)+\bigg(\Gamma_{0}^{(r)}\tanh(\frac{\beta\Omega_{0}}{2})-\frac{\Delta_{\rm eff}\Gamma_{0}}{\Omega_{0}}\bigg)\sin(\Omega_{0}t)
\bigg]
\right\}.
\eeq
In the above equations we have introduced the renormalized characteristic energy
\be
\label{eq:omega_x}
\Omega_{0}=\sqrt{\Delta_{{\rm eff}}^2+A^2}~,
\ee
where
\be
\label{eq:deltaeff_x}
\Delta_{{\rm eff}}=\Delta \left(\frac{\Delta}{\omega_c}\right)^{\alpha/(1-\alpha)}
\left[\Gamma(1-2\alpha)\cos(\pi\alpha)\right]^{1/[2(1-\alpha)]},
\ee
is the QB level splitting renormalized by dissipation with respect to the bare one $\Delta$.
This renormalization of the level splitting  it is the analogue of the Lamb shift~\cite{uli_book, gramich_2011, gramich_2014}, and it is due to the presence of weak dissipation.
Notice that the renormalized energy is always $\Omega_{0}\leq \Omega$, with $\Omega$ the bare characteristic energy of the QB in Equation~(\ref{eq:rabi}).
 Dissipation effects are reflected in the incoherent relaxation rate $\Gamma^{(r)}_{0}$ and dephasing rate $\Gamma_{0}$, which are given by
\beq
\label{eq:rates_x}
\Gamma_{0}^{(r)} &=& \frac{\pi \alpha \Delta_{\rm eff}^2}{\Omega_{0}}\coth[\beta \Omega_{0}/2]\nonumber\\
\Gamma_{0}&=&\frac{\Gamma_{0}^{(r)}}{2} + 2\pi \alpha\frac{A^2}{\beta\Omega^2_{0}} ~.
\eeq

The case of pure dephasing is described by setting $\theta=\pi$ in Equation~(\ref{eq:hint}), with the reservoir  coupled to $\sigma_z$, 
which induces a different dissipative dynamics.
Indeed, in this case the average energy associated to the QB can be written as
\beq
\label{eq:eqb_z}
&&\hspace{-1.6cm}\langle E_{{\rm QB}}(t)\rangle = \frac{\Delta}{2} \left\{1-\frac{\Delta}{\Omega_{\pi}}\tanh(\frac{\beta\Omega_{\pi}}{2})+\frac{\Delta}{\Omega_{\pi}}\bigg(\tanh(\frac{\beta\Omega_{\pi}}{2})-\frac{\Delta}{\Omega_{\pi}}\bigg)e^{-\Gamma_{\pi}^{(r)} t}+\right.\nonumber 
\\
&&\hspace{-1.6cm}+\left.\frac{e^{-\Gamma_{\pi} t}}{\Omega_{\pi}^2}\bigg[-A_{\rm eff}^2\cos(\Omega_{\pi}t)+\bigg(\Delta\Gamma_{\pi}^{(r)}\tanh(\frac{\beta\Omega_{\pi}}{2}) -\frac{\Delta^2\Gamma^{(r)}_{\pi}+A_{\rm eff}^2\Gamma_{\pi}}{\Omega_{\pi}}\bigg)\sin(\Omega_{\pi}t)
\bigg]\!\right\}
\eeq
and the charger contribution is given by
\beq
\label{eq:ec_z}
&&\hspace{-1.5cm}\langle E_{{\rm C}}(t)\rangle =\frac{AA_{\rm eff}}{2\Omega_{\pi}}\bigg\{-\tanh(\frac{\beta\Omega_{\pi}}{2})+\bigg[\tanh(\frac{\beta\Omega_{\pi}}{2})-\frac{\Delta}{\Omega_{\pi}}\bigg]e^{-\Gamma_{\pi}^{(r)} t}+\nonumber 
\\
&&\hspace{-1.5cm}+\left.\frac{e^{-\Gamma_{\pi} t}}{\Omega_{\pi}}\bigg[\Delta\cos(\Omega_{\pi}t)+\bigg(\Gamma_{\pi}^{(r)}\tanh(\frac{\beta\Omega_{\pi}}{2})-\frac{\Delta (\Gamma^{(r)}_{\pi}-\Gamma_{\pi})}{\Omega_{\pi}}\bigg)\sin(\Omega_{\pi}t)
\bigg]
\right\}.
\eeq
Notice the presence of a different characteristic energy
\be
\label{eq:omega_z}
\Omega_{\pi}=\sqrt{\Delta^2+A^2_{{\rm eff}}}~,
\ee
where the bare amplitude $A$, felt by the QB, gets renormalized by the presence of dissipation as
\be
\label{eq:aeff}
A_{{\rm eff}}=A\left(\frac{A}{\omega_c}\right)^{\alpha/(1-\alpha)}\left[\Gamma(1-2\alpha)\cos(\pi\alpha)\right]^{1/[2(1-\alpha)]}~.
\ee
Also here, one has $\Omega_{\pi}\leq \Omega$.
The relaxation and dephasing rates responsible for the damped oscillations are given by
\beq
\label{eq:rates_z}
\Gamma_{\pi}^{(r)} &=& \frac{\pi \alpha A^2_{\rm eff}}{\Omega_{\pi}}\coth[\beta \Omega_{\pi}/2]\nonumber\\
\Gamma_{\pi}&=&\frac{\Gamma_{\pi}^{(r)}}{2} + 2\pi \alpha \frac{\Delta^2}{\beta \Omega^2_{\pi}}~.
\eeq\\
In passing, we comment on another  particular value of the angle $\theta$. Indeed, looking at the mapping with the SB model, it is possible to fix a particular angle such that the tunneling amplitude $\Delta_0$ in Equation (\ref{eq:delta0}) vanishes
\be
\Delta_0=0 \to \bar{\theta}=2\arctan\left(\frac{\Delta}{A}\right)~.
\ee
Here, the competition between the two linear dissipative couplings $\sigma_x$ and $\sigma_z$ lead to a very peculiar behaviour. Interference between longitudinal and transverse noise have been discussed in related context of quantum dissipative systems~\cite{schnirman_2002, thorwart_2015, palm_2018}, and are reflected in the following expressions for the incoherent relaxation and dephasing rates
\beq
\Gamma^{(r)}_{\bar{\theta}}&=&0\nonumber\\\Gamma_{\bar{\theta}}&=&\frac{2\pi\alpha}{\beta}
\eeq
Putting these values in the weak damping expressions of the SB (see Section~\ref{app:weak_sb}) we finally get
\beq
\label{eq:etheta}
\langle E_{\rm QB}(t)\rangle&=&\frac{\Delta}{2}\left[1-\frac{\Delta^2}{\Omega^2}-\frac{A^2}{\Omega^2}e^{-\Gamma_{\bar{\theta}} t}[\cos(\Omega t)-\frac{\Gamma_{\bar{\theta}}}{\Omega}\sin(\Omega t)]\right]\\
\langle E_{\rm C}(t)\rangle&=&-\frac{A^2\Delta}{2\Omega^2}\left[1- e^{-\Gamma_{\bar{\theta}} t}[\cos(\Omega t)-\frac{\Gamma_{\bar{\theta}}}{\Omega}\sin(\Omega t)]\right].
\eeq
Interestingly, at zero temperature also $\Gamma_{\bar{\theta}}=0$ and the above expressions reduce to the same expression obtained for a closed QB under static driving (see Equation~(\ref{eq:closed})).
This peculiar choice of $\bar{\theta}$ thus lead to a very strong suppression of the effect caused by dissipation. However, to achieve this optimal point a precise control on the various parameters is required and other sources of dissipation (such as due to non linear couplings or $1/f$ noise) can become relevant~\cite{schnirman_2002, vion_2002, makhlin_2004, thorwart_2015, makhlin_2000, shen_2018}.

\subsection{Effect of dissipation on average energy}
\begin{figure}[ht]
\centering
\includegraphics[width=0.49\textwidth]{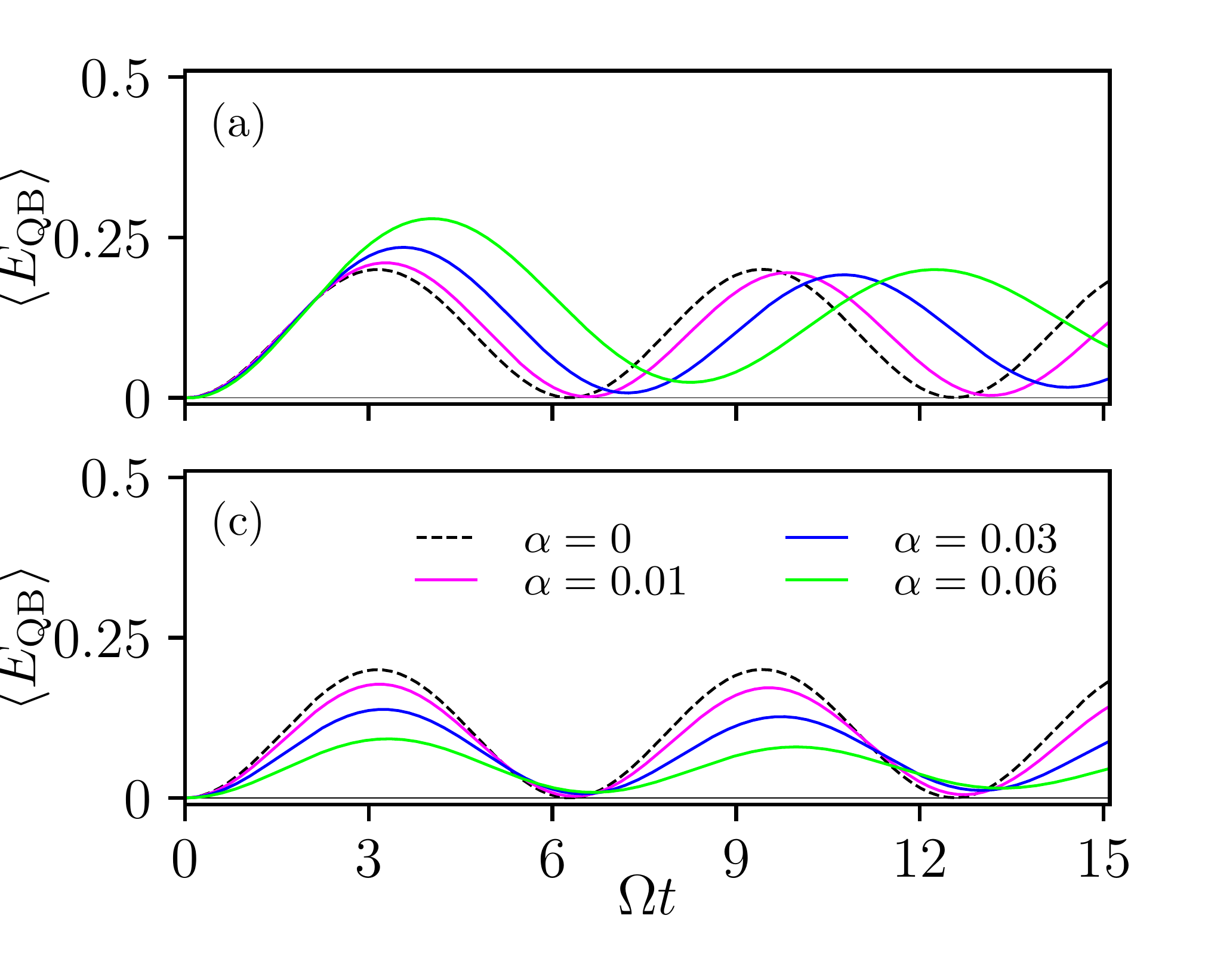}
\includegraphics[width=0.49\textwidth]{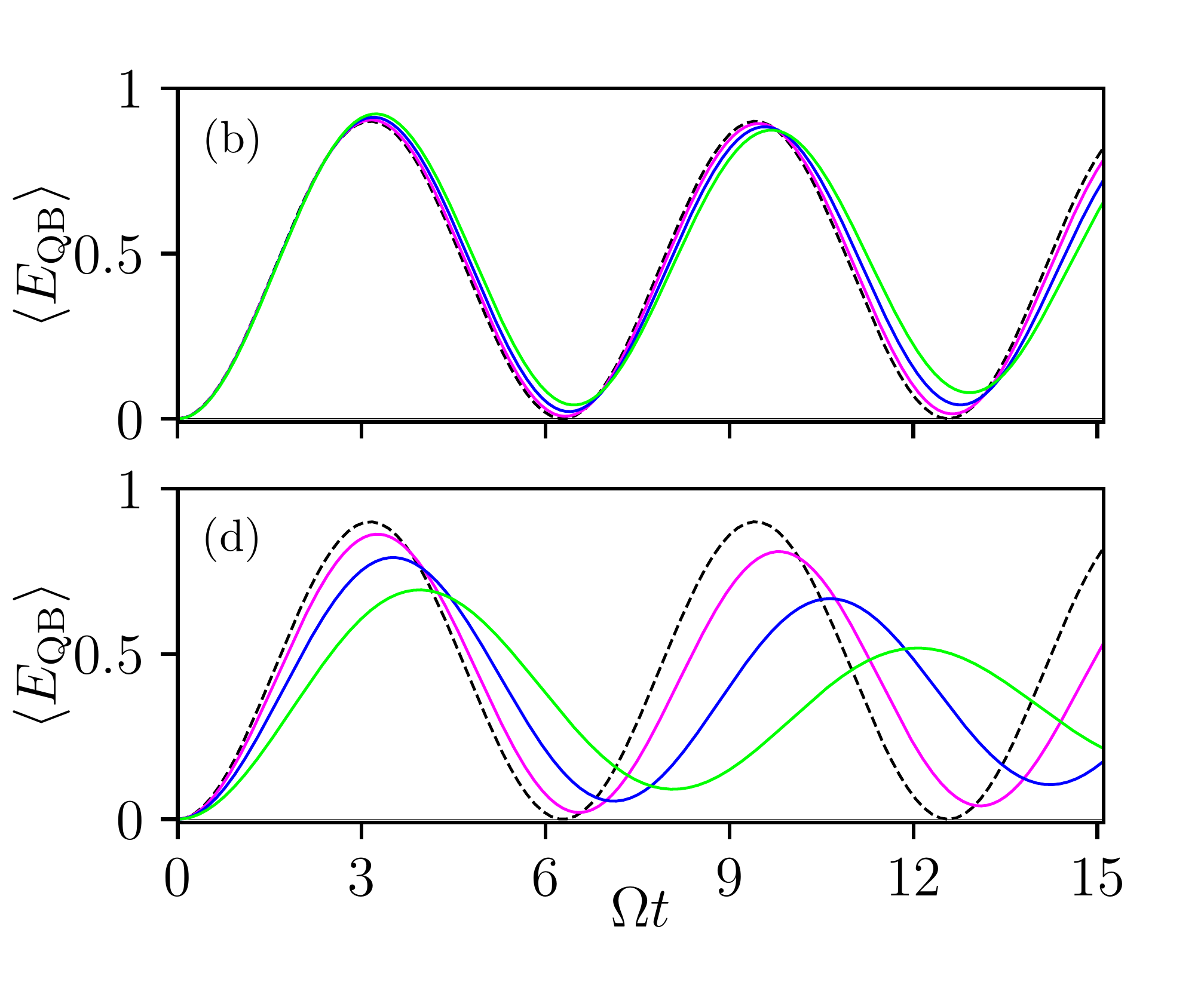}
	\caption{Time evolution of the average energy stored in the QB $\langle E_{{\rm QB}}\rangle$. Both decoherence coupling with $\theta=0$ (upper panels) and pure dephasing coupling with $\theta=\pi$ (lower panels) are shown as a function of $\Omega t$. Panels (a) and (c) refer to charging amplitude $A=0.5\Delta$, while panels (b) and (d) correspond to $A=3\Delta$. Different curves represent different dissipation strength $\alpha$. Other parameters are $\beta \Delta=10$, $\omega_c=500 \Delta$.}
 \label{fig1}
\end{figure}

We now discuss the results obtained in the previous section, showing how the effects of dissipation can modify the charging dynamics of a QB, in the weak damping regime at low temperature (setting $\beta\Delta=10$ in all the plots). 
In Figure~\ref{fig1} we show the time evolution of the average energy stored $\langle E_{{\rm QB}}(t)\rangle$ considering the decoherence coupling in Equation~(\ref{eq:eqb_x}) (see panels (a) and (b)) and pure dephasing one in Equation~(\ref{eq:eqb_z}) (see panels (c) and (d)).
Representative examples of driving amplitude $A=0.5\Delta$ (Figure~\ref{fig1}(a) and (c)) and $A=3\Delta$ (Figure~\ref{fig1}(b) and (d)) have been chosen.
The closed QB system driven by a static bias (see Equation~(\ref{eq:closed})) is also reported as a reference limit (see black dashed curves).
It is evident that, also in presence of dissipation, larger driving amplitude $A$ results in better charging of the QB (compare panels (b) and (d) versus panels (a) and (c)).
Overall, all curves present a damped oscillatory behaviour, whose amplitude is modulated by the exponential decay dictated by the incoherent relaxation and dephasing rates given in Equations~(\ref{eq:rates_x})-(\ref{eq:rates_z}). 
We recall that the rate expressions are different for the two dissipative couplings considered, and this gives rise to completely different relaxation dynamics as shown in the Figure.

The positions of maxima and minima are slightly shifted to higher values of $\Omega t$ with respect to the undamped case, effect particularly visible in panels (a) and panel (d). This can be understood by recalling that the characteristic energy gets renormalized with respect to the bare case $\Omega$ by the presence of dissipation as reported in Equations~(\ref{eq:omega_x})-(\ref{eq:omega_z}) (and is related to the Lamb shift).
Remarkably, looking at the first maximum (around $\Omega t\sim \pi$), an opposite behaviour between decoherence coupling $\theta=0$ and pure dephasing $\theta=\pi$ is observed while increasing the coupling strength $\alpha$.  Indeed, in the former case we observe an increase of the value of the maximum with respect to the closed system reference (black dashed curve). 

Again, this is a consequence of the renormalization of $\Delta$ in Equation~(\ref{eq:deltaeff_x}). Notice that this is true also for $A=3\Delta$, shown in panel (b), although it is less visible due to the larger value of the driving amplitude which partially mask the effect due to the renormalization of $\Delta_{{\rm eff}}$.
The opposite behaviour shown in panel (c)-(d) of Figure~\ref{fig1} can be explained by recalling the different mapping of the QB parameters in the pure dephasing case, where now the driving amplitude is renormalized to $A_{{\rm eff}}$, see Equation~(\ref{eq:aeff}).

\begin{figure}[ht]
\centering
\includegraphics[width=0.55\textwidth]{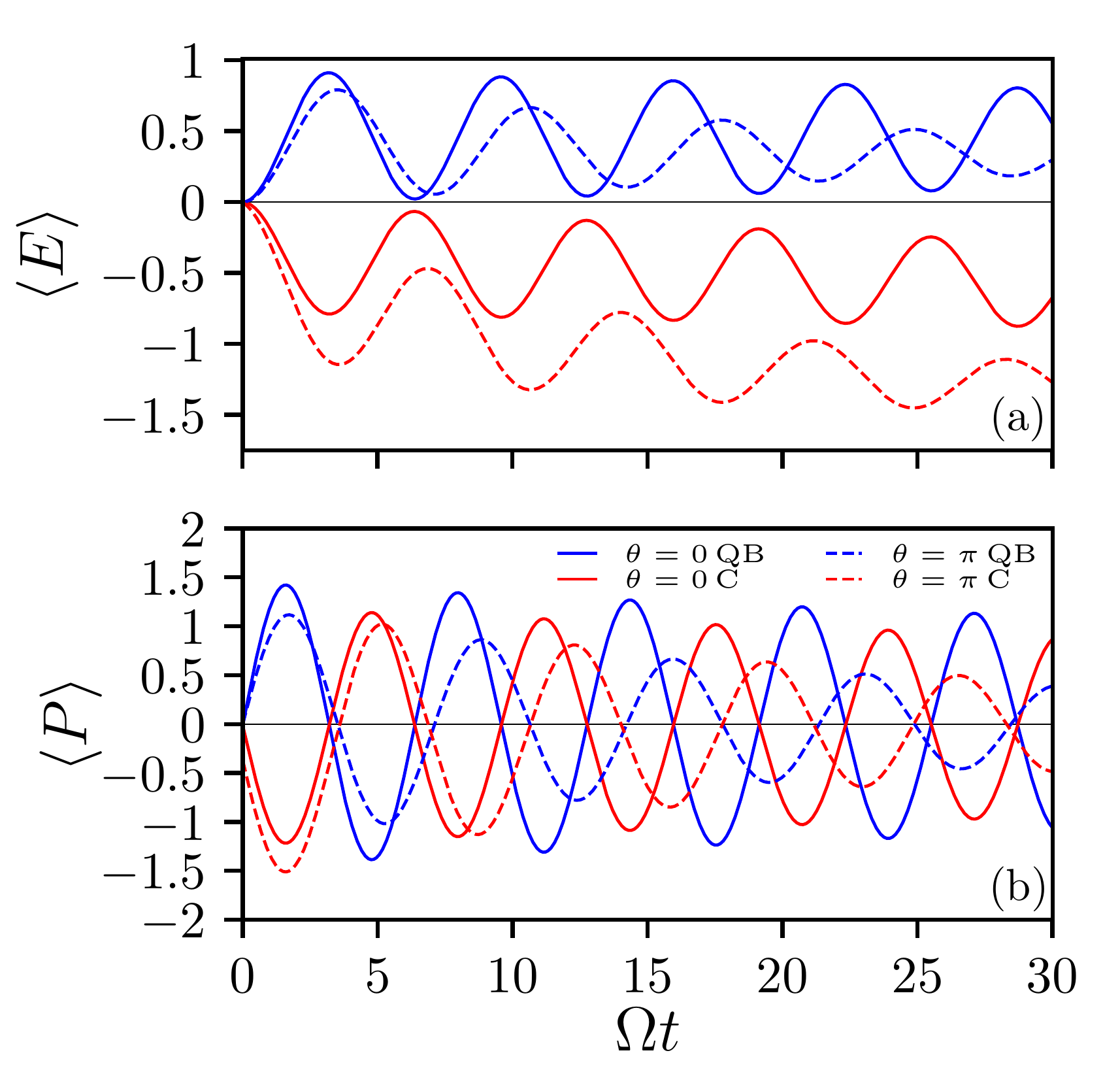}
\includegraphics[width=0.38\textwidth]{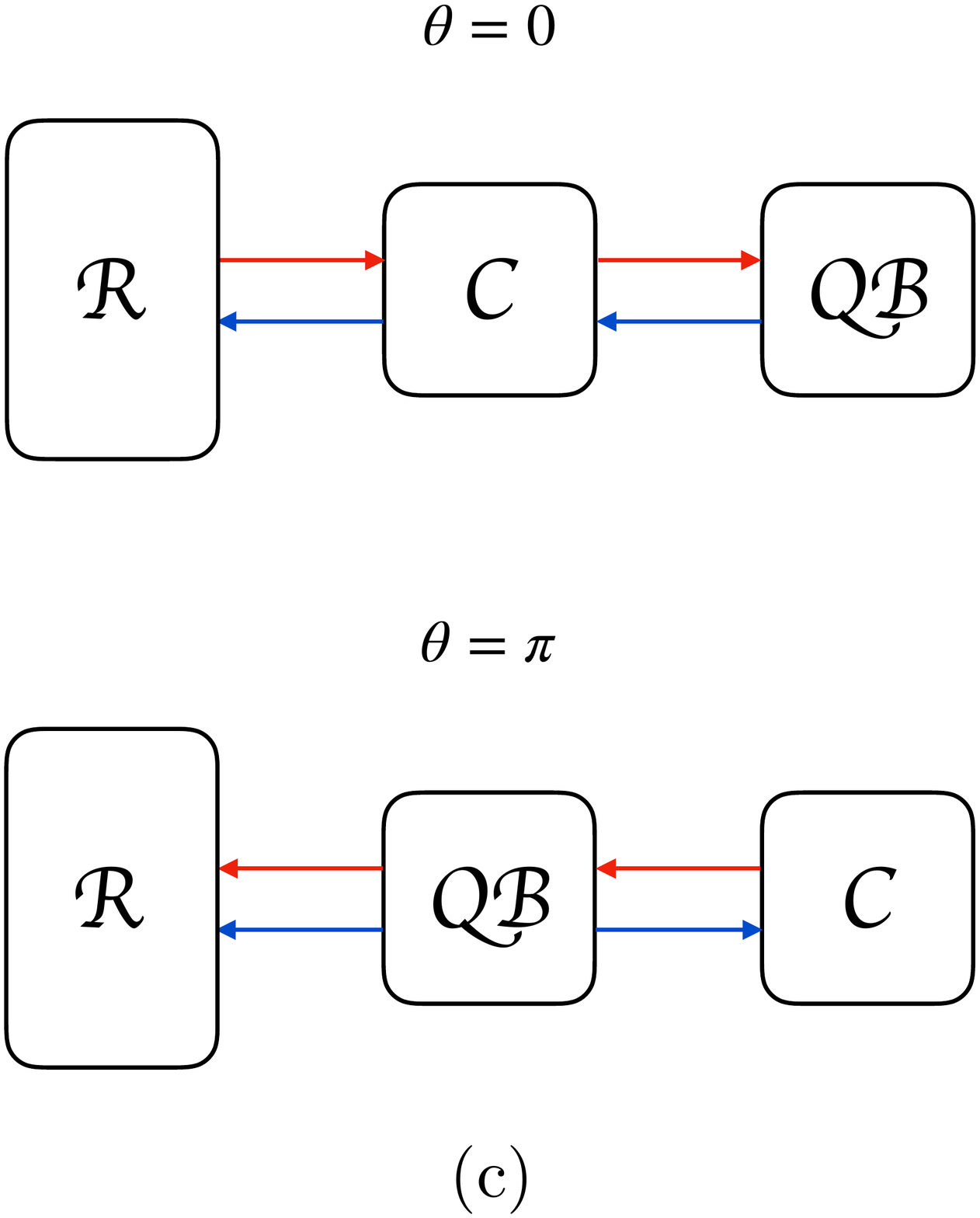}
	\caption{Time evolution of the average energy variation (a) and the corresponding power (b) associated to the QB (blue curves) and to the charger (red curves). Solid and dashed lines correspond to the case of decoherence ($\theta=0$) and pure dephasing ($\theta=\pi$) interaction with the reservoir, respectively. Dissipation strength has been fixed to $\alpha=0.03$ and driving amplitude to $A=3\Delta $. Other parameters as in Figure~\ref{fig1}. In panel (c) pictorial representation of the energy flow during charging and discharging process for the two couplings with the reservoir. }
\label{fig2}
\end{figure}

Direct comparison between the two dissipative coupling schemes is presented in Figure~\ref{fig2} (a)-(b) for a fixed driving amplitude $A=3\Delta$ and dissipation strength $\alpha=0.03$.
The plot reports (see blue lines) the average energy stored in the QB as a function of time, showing that in the case $\theta=0$ it is possible to achieve better charging of the QB (higher values of the maxima). 
Due to the different form of the rates in Equations~(\ref{eq:rates_x})-(\ref{eq:rates_z}) decoherence coupling is less affected by dissipation. Indeed, for the chosen parameters of Figure~\ref{fig2} (a) and (b) the corresponding rates are $\Gamma^{(r)}_{0}=0.021\Delta$, $\Gamma_{0}=0.028\Delta$ and $\Gamma^{(r)}_{\pi}=0.23\Delta$, $\Gamma_{\pi}=0.12\Delta$, respectively. This is a general trend, namely pure dephasing coupling is less efficient for charging process. Indeed, as discussed in related context of quantum dissipative systems~\cite{schnirman_2002, thorwart_2015, palm_2018, shen_2018}, pure dephasing coupling induced dynamics can be  explained effectively with a larger value of the dissipative coupling strength $\alpha$.

This relaxation process leads to a damping of the oscillations (see in Figure~\ref{fig2} (a)) resulting in a lower amount of energy that can be stored in the QB (quantified as the difference between a given maximum of $\langle E_{{\rm QB}}(t)\rangle$ and its preceding minimum). This is a general trend since dissipation induces relaxation dynamics toward a steady thermal state, whose values in the two discussed cases read
\beq
&&\langle E_{{\rm QB},\theta=0}(t\to\infty)\rangle = \frac{\Delta}{2} \left[1-\frac{\Delta_{\rm eff}}{\Omega_{0}}\tanh(\frac{\beta\Omega_{0}}{2})\right]\nonumber\\
&&\langle E_{{\rm QB},\theta=\pi}(t\to\infty)\rangle = \frac{\Delta}{2} \left[1-\frac{\Delta}{\Omega_{\pi}}\tanh(\frac{\beta\Omega_{\pi}}{2})\right]~.
\eeq
Being a thermal one, this steady state is equivalent to a passive state for the QB from which it is not possible to extract energy~\cite{Binder15, Campaioli17}. 
In Figure~\ref{fig2} (a) it is also reported (see red curves) the variation of energy associated to the charger $\langle E_{{\rm C}}(t)\rangle$ (see Equations~(\ref{eq:ec_x})-(\ref{eq:ec_z})).
Notice that curves associated to the QB and to the charger are not specular (with respect to the $x$ axis), reflecting the fact that a given amount of energy is also dissipated into the reservoir. This can be better visualized looking at the corresponding powers $\langle P_s(t)\rangle=\langle \dot{E}_s(t)\rangle$, shown in Figure \ref{fig2} (b). There, $\langle P_s\rangle >0$ indicates energy absorbed in the $s$-th channel ($s={\rm QB}, {\rm C}, {\rm RI}$), while $\langle P_s\rangle <0$ corresponds to energy released to the other subparts.

The marked differences between the solid blue ($\theta=0$) and dashed red ($\theta=\pi$) curves underlines the different charging (and discharging) process in the two cases, with a much better performance for the decoherence coupling  case. 

Indeed for $\theta=0$, the amplitude of first positive maximum $\langle P_{{\rm QB}}\rangle >0 $ corresponds almost to the first negative minimum $\langle P_{{\rm C}}\rangle <0 $ (energy released from the charger to the QB). Conversely, for $\theta=\pi$ (dashed lines) one has lower values for $\langle P_{{\rm QB}}\rangle$ and its value is also different from the corresponding $-\langle P_{{\rm C}}\rangle$, indicating that  part of the energy is already transfered to the bath.
Physically, this is due to the different energy flow (during charging and discharging dynamics) associated to the two different dissipative couplings.
These can be visualized as sketched in Figure~\ref{fig2} (c), where red arrows indicate charging process, while blue arrows refer to the reverse process. Notice that the difference between the two pictures in Figure~\ref{fig2} (c) reflects the different spin operator whose reservoir is directly coupled to (see Equation~(\ref{eq:hint})).

\begin{figure}[ht]
\centering
\includegraphics[width=0.478\textwidth]{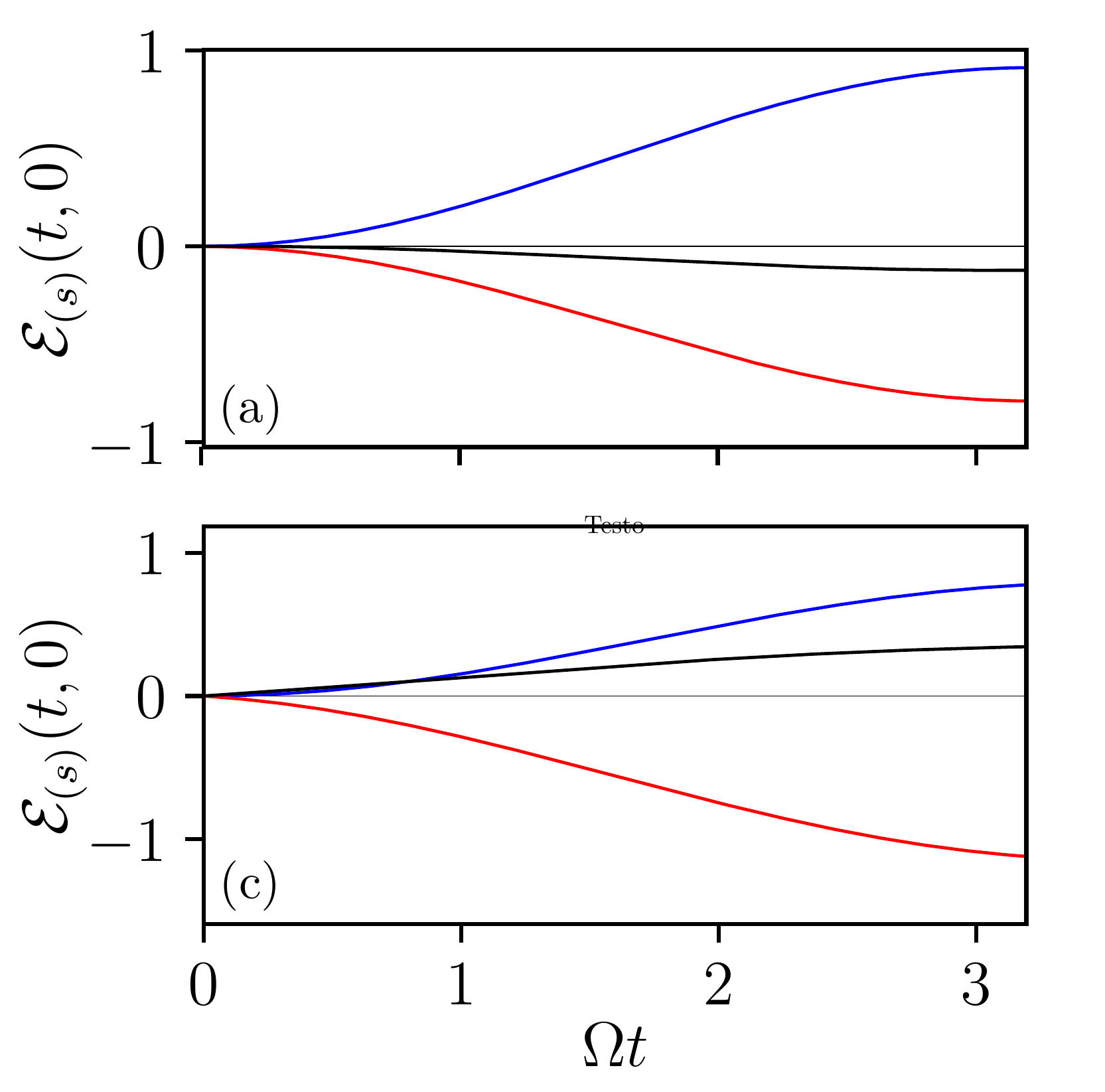}
\includegraphics[width=0.498\textwidth]{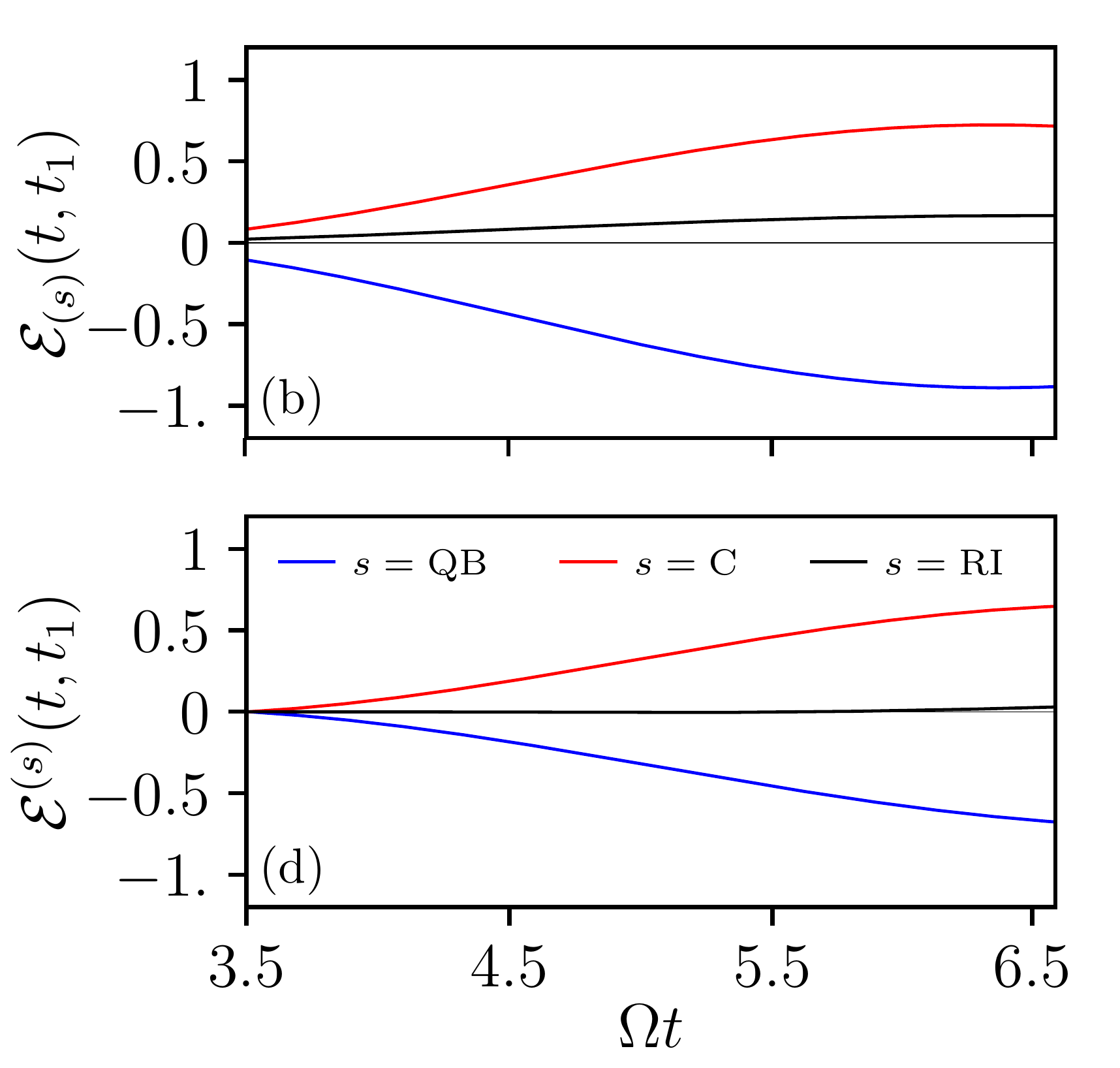}
\caption{Average energy flows in the first charging (panel (a) and (c)) and in the first discharging (panel (b) and (d)) windows. Upper panels refer to decoherence coupling ($\theta=0$) and lower panels refer to pure dephasing ($\theta=\pi$). Different curves indicate the average energy variation in the various channels during the charging/discharging process. Dissipation strength is $\alpha=0.03$ and driving amplitude $A=3\Delta$. Other parameters as in Figure~\ref{fig1}.}
 \label{fig3}
\end{figure}

To better understand the arrows directions in the different coupling mechanisms, we consider the first charging and the first discharging process and how energy is transferred between the various channels (QB, C, or RI) during these time windows. Being $t_1$ and $t_2$ the value of the first maxima and the first minima, respectively, we can define the energy variation in the first charging process as
\be
{\cal E}_s(t,0)\equiv \langle E_s(t)\rangle-\langle E_s(0)\rangle \quad 0\leq t\leq t_1~,
\ee
and in the first discharging process as
\be
{\cal E}_s(t,t_1)\equiv \langle E_s(t)\rangle -\langle E_s(t_1)\rangle \quad t_1\leq t\leq t_2~,
\ee

with $s=\rm{QB},\rm{C},\rm{RI}$. These quantities are reported in Figure~\ref{fig3} for a representative coupling strength $\alpha=0.03$ and driving amplitude $A=3\Delta$, where upper panels show the case $\theta=0$ and lower panels the case $\theta=\pi$.

Figure~\ref{fig3} (a)-(c) consider the energy variations associated to the QB, the charger and the reservoir contributions in the first charging interval $0\leq t\leq t_1$. As one can see looking at Figure~\ref{fig3} (a) the QB absorbs energy (positive energy variation) while both  the charger and the reservoir release energy (negative energy variation) in this time interval. 
In Figure~\ref{fig3} (c), instead, the charger supplies energy (negative energy variation) both to the QB and the reservoir (positive energy variation), i.e. a certain amount of energy is already dissipated into the thermal bath, thus reducing the amount of energy stored in the QB. These two situations are represented by the red arrows in the sketch of Figure~\ref{fig2} (c).
Figure~\ref{fig3} (b) and (d) report the first discharging process associated to decoherence (b) and pure dephasing (d) couplings, consistent with the blue arrows of Figure~\ref{fig2} (c). Here, in both cases energy is transferred from the QB (negative energy variation) to both the charger and the reservoir (positive energy variation). However, the precise amount of energy transferred to different channels is different in the two cases, as one can see looking at the red and black curves in Figure~\ref{fig3} (b) and (d).
This stress once more that not only the value of dissipation coupling strength $\alpha$ affects the charging dynamics, but also how the reservoir is coupled to the quantum system is crucial to determine both charging and discharging process of a QB. All the results confirm that the decoherence  coupling has a more efficient charging dynamics with respect to the pure dephasing case. Analogous conclusions between the two linear dissipative couplings can be obtained for higher values of dissipation strength.
Indeed, higher values of dissipation strengths have been also considered in the  NIBA framework and  reported in~\ref{app:niba}.

 Stronger dissipation strength would undermine charging performance of a QB, however the decoherence coupling channel remains the one that less influence QB dynamics.
Notice also that high temperature regime $\beta\Delta <1$ would lead to faster relaxations, and weaker charging performances, due to larger values of the associated rates (see Equations~(\ref{eq:rates_x})-(\ref{eq:rates_z}) and Figure~\ref{fig:niba}(b) in~\ref{app:niba} ).

\begin{figure}[ht]
\centering
\includegraphics[width=0.7\textwidth, height=0.45\textwidth]{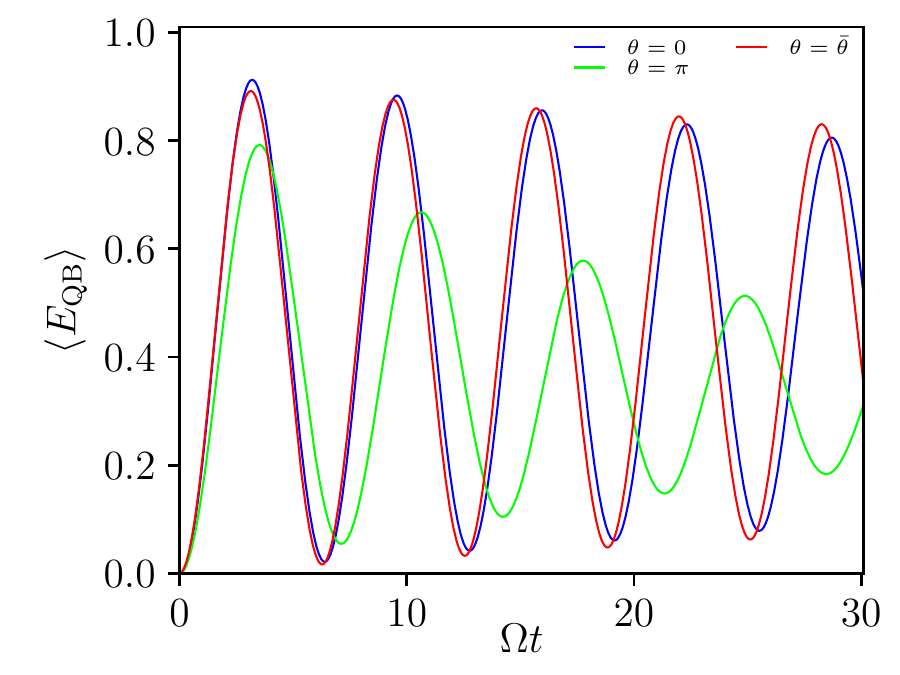}
\caption{Time evolution of the average energy stored in the QB. The plot shows comparison between decoherence coupling, pure dephasing and the optimal choice $\bar{\theta}$ for which linear coupling dissipation is strongly suppressed. Dissipative coupling strength is fixed to $\alpha=0.03$, $A=3\Delta$ and other parameters are the same as in Figure~\ref{fig1}.}
\label{figtheta}
\end{figure}

Before concluding this section, in Figure~\ref{figtheta} we show a comparison with the  case of the optimal choice which lead to a strong suppression of linear dissipation (see Equation~(\ref{eq:etheta})). 

As discussed in the previous section, by properly tuning the external parameters in presence of both longitudinal and transverse noise (with both decoherence and pure dephasing mechanisms) linear dissipation can be strongly suppressed. This results in a very slow relaxation, especially at long times, leading to very stable values of the maximum amount of average energy stored even after many cycles $\omega t$. However, it is worth to notice that considering the short time dynamics (confining to the first maxima shown in the Figure~\ref{figtheta}), again decoherence coupling mechanism lead to better charging, due to the renormalization effect induced by weak dissipation, reminiscent of the Lamb shift phenomena~\cite{schnirman_2002, gramich_2011, gramich_2014}.
Moreover, the choice of the optimal regime requires a very accurate fine tuning of the external parameters.
\subsection{Charging stability}
\label{sec:maintenance}
 \begin{figure}[ht]
\centering
\includegraphics[width=0.6\textwidth]{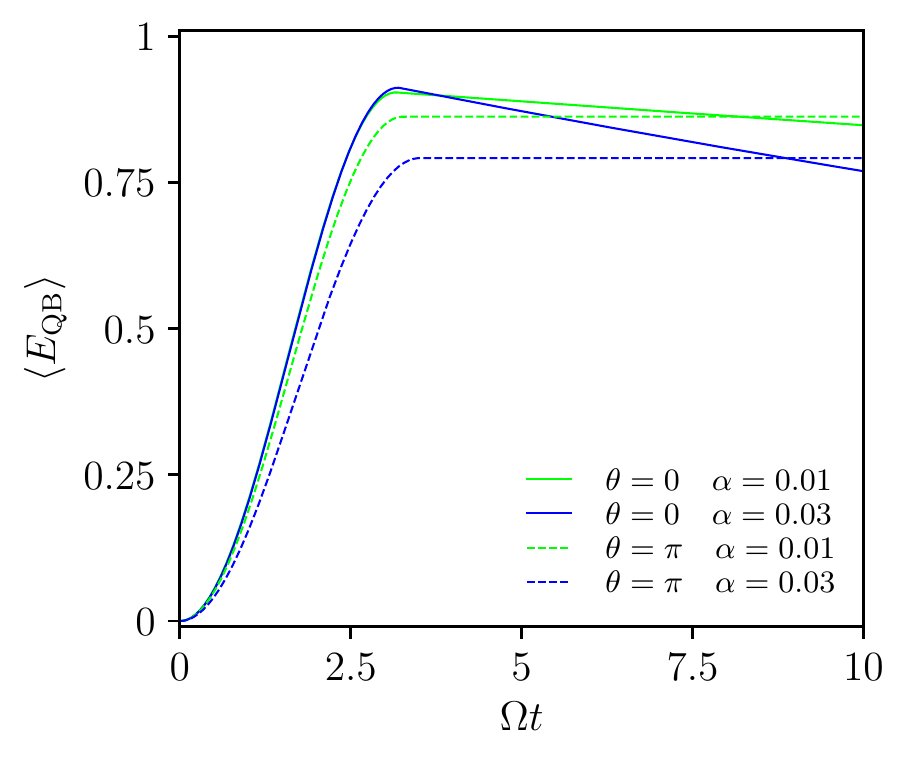}
\caption{Stability of average energy in presence of dissipation. Different curves represent different dissipation strengths and reservoir coupling (solid versus dashed curves). Note that the pure dephasing coupling ($\theta=\pi$) corresponds to  the QB effectively decoupled from the bath when the driving field is switched off. Other parameters are $A=3\Delta$, $\beta=10\Delta$ and $\omega_c=500\Delta$.}
\label{figmaintenance}
\end{figure}

Another important question regards how long a QB can retain a given amount of energy that has been stored during a charging process, in presence of dissipation. To answer this question, the following protocol is inspected. At $t\geq 0$ the QB is charged by interacting with the external charger with amplitude $A$ until a time $t_c>0$. Then, the charger is switched off, $A=0$, for $t>t_c$.
Of course, if one considers a closed system, the amount of energy stored in the QB until the time $t_c$ will remain constant at later times. However, in a more realistic situation the QB will dissipate part of its energy into the thermal bath.
It is therefore important to quantify the amount of energy retained in the QB, after the charging field has been switched off. To model this protocol, the QB is initially prepared at $t=0$ in a given state described by Equation~(\ref{eq:rhoqb}), e.g. starting from the ground state as considered before. The system then evolves under the action of the charger  with amplitude $A$ until the time $t_c$. At $t=t_c$ the reduced density matrix of the QB is
\beq
\rho(t_c)=\frac{1}{2}\left(\begin{array}{cc}
1+\langle \sigma_z(t_c)\rangle & \langle \sigma_x(t_c)\rangle - i \langle \sigma_y(t_c)\rangle \\
\langle \sigma_x(t_c)\rangle + i \langle \sigma_y(t_c)\rangle & 1-\langle \sigma_z(t_c)\rangle 
\end{array}\right)~.
\eeq
At times $t\geq t_c$ the amplitude $A$ is switched off and therefore it produces a different dynamics.
In the weak damping regime, this can be described in terms of  an effective initial reduced density matrix $\rho(t_c)$.

Due to the different coupling with the thermal bath for $t>t_c$ the dynamics of the QB will be still dissipative for a decoherence coupling, while  the case of pure dephasing results in total decoupling between the QB and the reservoir, once the driving field $A$ is switched off.
The charging storage will be then affected in a different way. Figure~\ref{figmaintenance} shows the time evolution of the average energy stored in the QB under this protocol (switching-off $A=0$ at time$t_c=t_1$).
Different curves refer to different coupling strengths $\alpha$. As one can see looking at the solid curves (corresponding to $\theta=0$), after the first maximum (at $t_c=t_1$) the average energy stored in the QB start to decrease. As mentioned above, the flat behaviour of the pure dephasing case is due to the effective decoupling of the QB from the bath (resembling the dynamics of a closed system). Interestingly, the decoherence coupling, for sufficiently weak dissipation strength, retains larger values of energy stored in the QB with respect to the other case in a wide time window. This confirm once more that decoherence coupling represents a good choice for the reservoir-quantum system engineering, resulting in both high values of average energy stored and both in terms of charging stability.  
 
\section{Conclusions}
\label{sec:conclusions}
In this work we have analized the dynamics of a single cell quantum
battery, modelled as a quantum two level system, charged by a classical
drive and coupled with an external thermal bath. We have focussed on the impact of dissipation on the charging performance of a QB, considering generic linear dissipative couplings with the environment, including both decoherence and pure dephasing mechanisms.
Exploiting a mapping to the spin-boson model and relying on a systematic perturbative expansion, we have found analytical expressions for the charging dynamics in the weak damping regime.
Here, charging and discharging dynamics
at short times have been studied in details, showing that decoherence
coupling between reservoir and QB results in better charging performance
with respect to pure dephasing one. Indeed, in the former case, the QB
achieves higher values of average energy stored, and its value is quite stable
also after switching-off the charging protocol for sufficiently weak coupling.
These findings, although based on a
simple QB model, represent important hints for a realistic
implementation of a QB in a solid state device, where the unavoidable
presence of an external environment has to be properly considered and
possibly engineered.

\section*{Acknowledgments}
M.C. acknowledges support from the Quant-EraNet project ``Supertop".

\appendix 
\section{Details on the unitary rotation}
\label{app:a}

To obtain Equation (\ref{widetildeH}) we perform a unitary rotation $\mathcal{R}$ on the Hamiltonian (\ref{eq:htot}) with 
\beq 
\mathcal{R}=e^{-i\phi\sigma_y}=(\cos(\phi) )\mathbf{1}-i (\sin(\phi)) \sigma_y =\left(\begin{array}{cc}
\cos(\phi) & -\sin(\phi) \\
\sin(\phi) & \cos(\phi)
\end{array}\right).
\eeq
Here, $\phi$ is a phase factor whose value has to be chosen in order to project the interaction part $H_{\rm I}$ only along the $z$ axis.
We thus consider the rotated interaction Hamiltonian
\be
\hspace{-1cm}\widetilde{H}_{{\rm I}}= \mathcal{R} H_{\rm I} \mathcal{R}^{-1} = \frac{1}{2}\left[\sin\bigg(\frac{\theta}{2}-2\phi\bigg)\sigma_z+\cos\bigg(2\phi-\frac{\theta}{2}\bigg)\sigma_x\right]\cdot\sum_{j}\lambda_j\left(a^{\dagger}_j+a_j\right).
\ee
In order to eliminate the contribution proportional to $\sigma_x$, we fix the value $\phi=(\theta+\pi)/4$, obtaining
\be 
\widetilde{H}_{\rm I}= -\frac{1}{2} \sigma_z\sum_{j}\lambda_j\left(a^{\dagger}_j+a_j\right).
\ee
Performing the same unitary rotation on the QB Hamiltonian one has
\be
\widetilde{H}_{{\rm QB}}=\mathcal{R} H_{{\rm QB}} \mathcal{R}^{-1} =\frac{\Delta}{2}\left[\sigma_z\cos(2\phi) +\sigma_x\sin(2\phi)\right],
\ee
which reduces to
\be 
\widetilde{H}_{\rm QB}= -\frac{\Delta}{2} \left[\sigma_z\sin(\theta/2) - \sigma_x\cos(\theta/2)\right]~, 
\ee
when $\phi=(\theta + \pi)/4$.
Finally, the term $H_{\rm C}$ related to the charger is transformed into
\be
\widetilde{H}_{{\rm C}}=\mathcal{R}H_{{\rm C}}\mathcal{R}^{-1}= \frac{A}{2}\left[-\sigma_z\sin(2\phi)+\sigma_x\cos(2\phi)\right],
\ee
and fixing $\phi$ it becomes
\be 
\widetilde{H}_{\rm C}= -\frac{A}{2}\left[\sigma_z\cos(\theta/2)+\sigma_x\sin(\theta/2)\right]. \ee
Of course, the term $H_{\rm R}$ is invariant under the rotation $\mathcal{R}$. We have therefore obtained Equation~(\ref{widetildeH}).
By applying the same rotation $\mathcal{R}$ on the density matrix of Equation (\ref{eq:rhoqb}) it is possible to obtain Equation~(\ref{initial}). Taking advantage of the property $p_R+p_L=1$ we obtain
\beq
\hspace{-2.3cm}\rho_{\rm SB}(0)&\equiv&\widetilde{\rho}(0)=\mathcal{R}\rho(0)\mathcal{R}^{-1}=\nonumber \\
\hspace{-2.3cm}&=&\left(\!\!\begin{array}{cc}
p_R\cos(2\phi) +\frac{1-\cos(2\phi)}{2}-a\sin(2\phi) & \frac{p_R-p_L}{2}\sin(2\phi)+a\cos(2\phi)-ib  \\
\frac{p_R-p_L}{2}\sin(2\phi)+a\cos(2\phi)+ib & p_L\cos(2\phi) +\frac{1-\cos(2\phi)}{2}+a\sin(2\phi)
\end{array}\!\!\right)
\eeq
where we have identified the reduced density matrix of the SB in Equation~(\ref{initial}) with
\be\begin{array}{rcl}
\bar{p}_R&=&-p_R\sin\frac{\theta}{2}+\frac{1+\sin\frac{\theta}{2}}{2}-a\cos\frac{\theta}{2}\\\\
\bar{p}_L&=&-p_L\sin\frac{\theta}{2}+\frac{1+\sin\frac{\theta}{2}}{2}+a\cos\frac{\theta}{2}\\\\
\bar{a}&=&-a\sin\frac{\theta}{2}+\frac{p_R-p_L}{2}\cos\frac{\theta}{2}\\\\
\bar{b}&=&b,
\end{array}\ee
which coincide with Equations (\ref{pr})-(\ref{b}) of the main text.

\section{Dissipative Kernels}
\label{app:kernels}
This Appendix details the results introduced in  Subsection \ref{GME}. We will  derive the general expressions for $\langle\sigma_{z/x}(t)\rangle$ in the SB model,  discussing the dissipative kernels $K_{1/2,z}^{(\pm)}(\tau)$ and  $K_{1/2,x}^{(\pm)}(\tau)$,  which are the main building blocks to construct the generalized master equation. The useful links introduced in Equations (\ref{generallinks1}) and (\ref{generallinks2}) will be demonstrated. 

The dynamics of $\langle\sigma_{k}(t)\rangle$ ($k=x,y,z$) can be studied evaluating the time evolution of the reduced density matrix, starting from the initial condition at $t=0$. We have indeed 
\beq
\langle\sigma_z(t)\rangle &=& \rho_{1,1}(t)-\rho_{-1,-1}(t),\nonumber\\
\langle\sigma_x(t)\rangle&=& \rho_{1,-1}(t)+\rho_{-1,1}(t),\nonumber\\
\langle\sigma_y(t)\rangle&=& i[\rho_{1,-1}(t)-\rho_{-1,1}(t)]
\label{Brho}
\eeq
where $\rho_{1,1}(t)$ and $\rho_{-1,-1}(t)$ are the populations, while $\rho_{1,-1}(t)$ and $\rho_{-1,1}(t)$ are the so-called coherence terms. 
As shown in details in References \cite{sassetti_1990,uli_book,grifoni_1996,grifoni_1999} the reduced density matrix can be expressed in terms of a real-time double path integral 
\be
\rho_{\sigma,\sigma'}(t)=\sum_{\sigma_0,\sigma_0'=\pm 1}\int{\cal D}\sigma\int{\cal D}\sigma' {\cal A}[\sigma] {\cal A}^*[\sigma']{\cal F}[\sigma,\sigma']\rho_{\sigma_0,\sigma_0'}(t=0).
\ee
Here,  the symbol $\int{\cal D}\sigma\int{\cal D}\sigma'$ means the summation over all spin paths which for a two-state system are 
$\sigma(t')=\pm 1$ and $\sigma'(t')=\pm 1$ with $0\le t'\le t$. The boundary conditions are
$\sigma(t)=\sigma$, $\sigma'(t)=\sigma'$ at final time $t$ and $\sigma(0)=\sigma_0$, $\sigma'(0)=\sigma'_0$ at inital time.
The elements of the initial reduced density matrix are $\rho_{\sigma_0,\sigma_0'}(0)$  and correspond to the matrix in Equation (\ref{initial})   
\beq
\rho(0)=\left(\begin{array}{cc}
\bar{p}_R & \bar{a}-i \bar{b} \\
\bar{a}+i \bar{b} & \bar{p}_L
\label{Binitial}
\end{array}\right).
\eeq 
The quantity ${\cal A}[\sigma]$ is the probability amplitude for the free (undamped) two-level system  to follow the path $\sigma(t')$. It consists of a contribution due to the tunneling processes which is $i\Delta_0/2$  for each  jump with spin changes $\sigma=\pm 1\to \mp 1$. In addition it receives a contribution from the static bias given by  $\exp\left\{{i\epsilon_0\int_0^{t}dt'\sigma(t')}\right\}$. The effects of the bath are included in the Feynman-Vernon influence function ${\cal F}[\sigma,\sigma']$ which is obtained after tracing out the thermal reservoir \cite{feynman_1963}
\be
\hspace{-2.5cm}{\cal F}[\sigma,\sigma']=\exp{\left\{-\frac{1}{4}\int_0^{t} dt'\int_0^{t'} dt''[\sigma(t')-\sigma'(t')][{\cal L}(t'-t'')\sigma(t'')- {\cal L}(t''-t')\sigma'(t'')]
\right\}}.
\label{Binflu}
\ee
Here, the Kernel ${\cal L}(t)$ represents the autocorrelation function of the fluctuating force which the reservoir exerts on the two-level system and it has the following spectral representation \cite{uli_book}
\be
{\cal L}(t)=\int_0^{\infty}d\omega
J(\omega)\left[\cos(\omega t)\coth(\beta\omega/2)-i\sin(\omega t)\right],
\label{forcecorr}
\ee
with $J(\omega)$ the spectral density of the bath given in Equation (\ref{ohmic}).  Notice that the presence of the bath induces  interactions among the paths at different times, with  a non-markovian time memory.
Since the spin paths are piecewise constant  with sudden jumps in between it is convenient to perform integration by parts in the influence functional (\ref{Binflu}) introducing the function 
$\ddot Q(t)={\cal L}(t)$. Below we quote the result avoiding to explicitly writing the terms containing the boundaries of $\sigma(t')$ and $\sigma'(t')$ in $t'=t,0$
\beq
\hspace{-1cm}{\cal F}[\sigma,\sigma']&=&\exp\Big\{\frac{1}{4}\int_0^{t} dt'\int_0^{t'} dt''
[\dot\sigma(t')-\dot\sigma'(t')]Q'(t'-t'')[\dot\sigma(t'')-\dot\sigma'(t'')]\nonumber\\
&+&i[\dot\sigma(t')-\dot\sigma'(t')]Q''(t'-t'')[\dot\sigma(t'')+\dot\sigma'(t'')]\Big\},
\label{Binflu1}
\eeq

where
\be
\hspace{-2cm}Q(t)=Q'(t)+iQ''(t)=\int_0^{\infty}d\omega
\frac{J(\omega)}{\omega^2}\Big[(1-\cos(\omega t))\coth(\beta\omega/2)+i\sin(\omega t)\Big].
\label{forcecorrq}
\ee
This dissipative correlator  can be evaluated for the Ohmic spectral density (\ref{ohmic}) and it reads ($\beta\omega_c\gg 1$)
\begin{eqnarray}
Q'(t)&=&2\alpha\ln\left[\frac{\beta\sqrt{1+\omega_{{ c}}^2 t^2}}{\pi t}\sinh\left(\frac{\pi t}{\beta}\right)\right]\\
Q''(t)&=&2\alpha\arctan(\omega_{{ c}}t).
\label{eq:qgeneral}
\end{eqnarray}
Considering the scaling limit of large cut-off $\omega_c$, $Q(t)$ reduces to 
\be
Q'(t)=2\alpha\ln\left[\frac{\beta\omega_{{ c}}}{\pi}\sinh\left(\frac{\pi |t|}{\beta}\right)\right],\qquad
Q''(t)=\pi\alpha{\rm sgn}(t).
\label{Bscaling}
\ee
In the following we will consider this regime.
The time evolution of the density matrix will be now written in terms of the summations over all possible paths which consists in an expansion in the number of tunneling transitions $\Delta_0$, weighted by the different factors discussed above. The details of this procedure are discussed in several references, see { e.g.}   \cite{uli_book,grifoni_1996,grifoni_1999}. For this reason, we decided to write and comment  the series expansion for   $\langle\sigma_z(t)\rangle$, which will allow us to demonstrate the important relation introduced in Equation (\ref{generallinks1}). We choose as  initial density matrix the one given in Equation (\ref{Binitial}). Inserting the different spin paths and considering the diagonal elements of $\rho(t)$ one obtains the following decomposition for $\langle\sigma_z(t)\rangle$ 
\be
\langle\sigma_z(t)\rangle=\langle\sigma_{z,0}(t)\rangle+\langle\sigma_{z,\bar{a}}(t)\rangle+\langle\sigma_{z,\bar{b}}(t)\rangle,
\label{Bdecomposition}
\ee
where \cite{grifoni_1996}
\beq
\langle\sigma_{z,0}(t)\rangle &=& (\bar p_R-\bar p_L)+\sum_{n=1}^{\infty}\Delta_0^{2n}\left(\frac{-1}{2}\right)^{n}\int_0^t dt_{2n}\cdots \int_0^{t_2} dt_{1}\cdot\nonumber\\ 
&&\cdot\sum_{\xi_1\cdots\xi_n=\pm 1}\left[(\bar p_R-\bar p_L){F}_{n,1}^{(+)}C_{n,1}^{(+)}-{F}_{n,1}^{(-)}C_{n,1}^{(-)}\right]
\label{Bsigmaz0}
\eeq
and \cite{grifoni_1999}
\beq
\hspace{-2cm}\langle\sigma_{z,\bar a}(t)\rangle &=&-2{\bar a} \sum_{n=1}^{\infty}\Delta_0^{2n-1}\left(\frac{-1}{2}\right)^{n}\int_0^t dt_{2n-1}\cdots \int_0^{t_2} dt_{1}\sum_{\xi_1\cdots\xi_n=\pm 1}\xi_1{F}_{n,2}C_{n,2}^{(-)}
\label{Bsigmaza}\\
\hspace{-2cm}\langle\sigma_{z,\bar b}(t)\rangle &=& 2{\bar b} \sum_{n=1}^{\infty}\Delta_0^{2n-1}\left(\frac{-1}{2}\right)^{n}\int_0^t dt_{2n-1}\cdots \int_0^{t_2} dt_{1}\sum_{\xi_1\cdots\xi_n=\pm 1}{F}_{n,2}C_{n,2}^{(+)}.
\label{Bsigmazb}
\eeq
Notice that the initial condition is $\langle\sigma_{z}(t=0)\rangle=\langle\sigma_{z,0}(t=0)\rangle=\bar p_R-\bar p_L$.
The integrations are over the flips times, related to the different transitions of a given spin path. These times are time-ordered with $0\le t_1\le\cdots\le t_{2n}\le t$. Notice that in $\langle\sigma_{z,\bar a}(t)\rangle$ and $\langle\sigma_{z,\bar b}(t)\rangle$ there is one  integration left with $t_0=0$. 
The sum $\sum_{\xi_j=\pm 1}$ is over the integer variables  $\xi_1\cdots\xi_{n}$ which can assume, each, the values $\pm 1$.
Let us now discuss the different terms inside the integrals. The  factors $C_{n,1/2}^{(\pm)}$ are the even and odd phase associated to the static bias term $\epsilon_0$
\beq
\hspace{-2cm}C_{n,1}^{(+)}&=&\cos[\epsilon_0\sum_{j=1}^n\xi_j(t_{2j}-t_{2j-1})]\qquad
C_{n,1}^{(-)}=\sin[\epsilon_0\sum_{j=1}^n\xi_j(t_{2j}-t_{2j-1})],\\
\label{Aphase}
\hspace{-2cm}C_{n,2}^{(+)}&=&\cos[\epsilon_0\sum_{j=1}^n\xi_j(t_{2j-1}-t_{2j-2})]\qquad
C_{n,2}^{(-)}=\sin[\epsilon_0\sum_{j=1}^n\xi_j(t_{2j-1}-t_{2j-2})].
\label{Bphase}
\eeq

The influence of the bath  is included in the functions ${F}_{n,1}^{(\pm)}$ and ${F}_{n,2}$, which correlate the spin's transitions at different times. In the scaling limit  they are expressed in compact notations,  using the expression (\ref{Bscaling}) for the interaction $Q(t)$. We have 
\beq
\label{eq:bf}
\hspace{-2cm}{F}_{n,1}^{(+)}(t_{2n},\cdots, t_1)&=&G_{n,1}\cdot\left[\cos(\pi\alpha)\right]^{n},\qquad {F}_{n,1}^{(-)}(t_{2n},\cdots, t_1)=\xi_1\tan(\pi\alpha){F}_{n,1}^{(+)}\\
\hspace{-2cm}{F}_{n,2}(t_{2n-1},\cdots, t_0)&=&G_{n,2}\cdot\left[\cos(\pi\alpha)\right]^{n-1},
\eeq
where
\beq
G_{n,1}(t_{2n},\cdots, t_1)&=&\exp\left( -\sum_{j=1}^{n}Q'_{{2j},{2j-1}}-\sum_{j=2}^{n}\sum_{k=1}^{j-1}\xi_j\xi_k \Lambda^{(1)}_{j,k}\right),
\label{eq:bg1}\\
G_{n,2}(t_{2n-1},\cdots, t_0)&=&\exp\left(-\sum_{j=0}^{n-1}Q'_{{2j+1},{2j}}-\sum_{j=2}^{n}\sum_{k=1}^{j-1}\xi_j\xi_k \Lambda^{(2)}_{j,k}\right).
\label{eq:bg2}
\eeq
Here, we defined $Q'_{{j},{k}}=Q'({t_{j}-t_{k}})$ and 
\beq
\Lambda^{(1)}_{j,k}=Q'_{{2j},{2k-1}}+Q'_{{2j-1},{2k}}-Q'_{{2j},{2k}}-Q'_{{2j-1},{2k-1}}\nonumber\\
\Lambda^{(2)}_{j,k}=Q'_{{2j-1},{2k-2}}+Q'_{{2j-2},{2k-1}}-Q'_{{2j-1},{2k-1}}-Q'_{{2j-2},{2k-2}}.
\eeq

We are now in the position to demonstrate the important links quoted in Equations (\ref{generallinks1})  between  $\langle\sigma_{z,0}(t)\rangle$ and  $\langle\sigma_{z,{\bar a}/{\bar b}}(t)\rangle$. For this, we write the even an odd part of $\langle\sigma_{z,0}(t)\rangle$ with respect to the bias $\epsilon_0$, $\langle\sigma_{z,0}(t)\rangle=\langle\sigma_{z,0}^{(+)}(t)\rangle+\langle\sigma_{z,0}^{(-)}(t)\rangle$,
whose series is implicitly defined in Equation (\ref{Bsigmaz0})
\beq
\hspace{-2.3cm}\langle\sigma_{z,0}^{(+)}(t)\rangle &=&\!(\bar p_R-\bar p_L)\!\left[1+\sum_{n=1}^{\infty}\Delta_0^{2n}\left(\frac{-1}{2}\right)^{n}\int_0^t dt_{2n}\cdots\int_0^{t_2} dt_{1}\sum_{\xi_1\cdots\xi_n=\pm 1}{F}_{n,1}^{(+)}C_{n,1}^{(+)}\right]\\
\hspace{-2.3cm}\langle\sigma_{z,0}^{(-)}(t)\rangle &=&-\sum_{n=1}^{\infty}\Delta_0^{2n}\left(\frac{-1}{2}\right)^{n}\int_0^t dt_{2n}\cdots \int_0^{t_2} dt_{1}\sum_{\xi_1\cdots\xi_n=\pm 1}{F}_{n,1}^{(-)}C_{n,1}^{(-)}.
\eeq
The time derivative of these two parts is easily performed by fixing the last integration time $t_{2n}=t$.
By comparing these expressions to the ones written in Equations (\ref{Bsigmaza}) and (\ref{Bsigmazb}) for  $\langle\sigma_{z,{\bar a}/{\bar b}}(t)\rangle$ we identify, after a proper change of integration variables, the links quoted in the main text
\begin{eqnarray}
\langle\sigma_{z,\bar{a}}(t)\rangle&=& \frac{2\bar{a}}{\Delta_0\sin(\pi\alpha)}\frac{d}{dt}\langle{\sigma}^{(-)}_{z,0}(t)\rangle\nonumber\\
\langle\sigma_{z,\bar{b}}(t)\rangle&=& \frac{2\bar{b}}{\Delta_0\cos(\pi\alpha)(\bar{p}_R-\bar{p}_L)}\frac{d}{dt}\langle{\sigma^{(+)}_{z,0}(t)}\rangle.
\label{Bgenerallinks1}
\end{eqnarray}
Notice that the first Equation is well defined for $\alpha\to 0$ since the quantity $\frac{d}{dt}\langle{\sigma^{(-)}_{z,0}(t)}\rangle$ is proportional to $\sin(\pi\alpha)$. 

We now comment on the possibility to write an exact master equation for $\langle\sigma_{z}(t)\rangle$ starting from its
series expression obtained above.
Although the series cannot be in general solved exactly, it is always possible to 
link the time derivative $d\langle\sigma_{z}(t)\rangle/dt$ with $\langle\sigma_{z}(t)\rangle$ itself. This is achieved via direct comparison, term by term,  of their series expansions.  Notice that  $d\langle\sigma_{z}(t)\rangle/dt$ has also a formal series expansions obtained directly by the time derivation of  $\langle\sigma_{z}(t)\rangle$. This procedure allows us to write, on a general ground, the following generalized master equation \cite{grifoni_1996,grifoni_1999,uli_book}
\be
\hspace{-2.3cm}\frac{d\langle \sigma_{z}(t)\rangle }{dt}= \int_0^t dt' [K^{(-)}_{1,z} (t-t') - K^{(+)}_{1,z}(t-t')\langle \sigma_{z}(t')\rangle ] +2\bar{a} K^{(-)}_{2,z}(t)-2\bar{b} K^{(+)}_{2,z}(t).
\label{Beq:gme_sigmaz}
\ee
Here, the kernels $K^{(\pm)}_{1/2,z}(t-t')$  are constructed by matching the iterative solution  represented in Equation (\ref{Beq:gme_sigmaz}) 
with the exact formal series for $\langle\sigma_{z}(t)\rangle$ in Equations (\ref{Bdecomposition})-(\ref{Bsigmazb}). They are expressed as series and they represent the irreducible components of the exact series for  $\langle\sigma_{z}(t)\rangle$ \cite{uli_book}.
Due to the links demonstrated in Equations (\ref{Bgenerallinks1}) it directly follows that the kernels associated to the off diagonal terms of the initial density matrix are related to the kernel $K^{(\pm)}_{1,z} (t)$ as
\begin{eqnarray}
 K_{2,z}^{(+)}(t)&=& \frac{1}{\Delta_0\cos(\pi\alpha)}K_{1,z}^{(+)}(t)\nonumber\\
 K_{2,z}^{(-)}(t)&=& \frac{1}{\Delta_0\sin(\pi\alpha)}K_{1,z}^{(-)}(t).
\label{linksK}
\end{eqnarray}
For this reason we quote below only the series expansions obtained for $K^{(\pm)}_{1,z} (t)$. We have ($t>t'$) \cite{grifoni_1996,uli_book}
\be
\hspace{-2.3cm}K^{(\pm)}_{1,z}(t-t')= {\cal K}^{(\pm)}_{1,z}(t-t')-\sum_{n=2}^{\infty}\Delta_0^{2n}\left(\frac{-1}{2}\right)^{n}\int_{t'}^t dt_{2n-1}\cdots \int_{t'}^{t_3} dt_{2}
\sum_{\xi_1\cdots\xi_n=\pm 1} {\widetilde{\cal F}}_{n,1}^{(\pm)}.
\label{Bkappa}
\ee
Here, the new functionals ${\widetilde{\cal F}}_{n,1}^{(\pm)}(t_{2n},t_{2n-1},\cdots t_2,t_1)$, depend on $2n$-times with the first and the last 
blocked at $t_1=t'$ $t_{2n}=t$. They correspond to the irreducible influence functionals with $2n$ number of transitions,
obtained from ${F}_{n,1}^{(\pm)}C_{n,1}^{(\pm)}$ by subtracting all possibilities of factorizing the influence functions into independent clusters (reducible components). Their expression is discussed in \cite{uli_book}, for a path with $2n$ transitions they are
\beq
\hspace{-2.3cm}{\widetilde{\cal F}}_{n,1}^{(+)}&=&{F}_{n,1}^{(+)}C_{n,1}^{(+)}- \sum_{j=2}^{n}(-1)^j\!\!\sum_{m_1,\cdots,m_j}\!\!{F}_{m_1,1}^{(+)}C_{m_1,1}^{(+)}\cdot{F}_{m_2,1}^{(+)}C_{m_2,1}^{(+)}\cdots {F}_{m_j,1}^{(+)}C_{m_j,1}^{(+)}\cdot
\delta_{m_1+\cdots+m_j,n}\nonumber\\
\hspace{-2.3cm}{\widetilde{\cal F}}_{n,1}^{(-)}&=&{F}_{n,1}^{(-)}C_{n,1}^{(-)}- \sum_{j=2}^{n}(-1)^j\!\!\sum_{m_1,\cdots,m_j}\!\!{F}_{m_1,1}^{(+)}C_{m_1,1}^{(+)}\cdot{F}_{m_2,1}^{(+)}C_{m_2,1}^{(+)}\cdots {F}_{m_j,1}^{(-)}C_{m_j,1}^{(-)}\cdot
\delta_{m_1+\cdots+m_j,n}.\nonumber\\
\hspace{-2.3cm}&&
\eeq
The inner sum is over positive integers $m_k$, and in the subtracting part the bath correlations are only inside each individual term ${F}_{m_k,1}^{(+)}C_{m_k,1}^{(+)}$ without correlations among them. Any terms has the time variables written with time growing from right to left. 
The first term ${\cal K}^{(\pm)}_{1,z}(t-t')$ in (\ref{Bkappa})  are the contributions at order  $\Delta_0^2$ (the series starts with $\Delta_0^4$) and are without internal integrations. For these terms the irreducible functional corresponds directly to the functional itself ${\widetilde{\cal F}}_{n=1,1}^{(\pm)}={F}_{n=1,1}^{(\pm)}C_{n=1,1}^{(\pm)}$. Their explicit expressions are
\beq
\label{eq:calk_za}
{\cal K}^{(+)}_{1,z} (t-t')&=& \Delta^2_0  e^{-Q'(t-t')}\cos(\pi\alpha)\cos[\epsilon_0(t-t')]\\
\label{eq:calk_zb}
{\cal K}^{(-)}_{1,z} (t-t')&=& \Delta^2_0  e^{-Q'(t-t')}\sin(\pi\alpha) \sin[\epsilon_0(t-t')].
\eeq

We conclude by commenting on the properties of $\langle\sigma_{x}(t)\rangle$. The $x$-component has also a series expansion in the tunneling amplitude $\Delta_0$, obtained similarly to what already  discussed  for the $z$-component, for this reason we will omit the details quoting directly the most important results. First,  $\langle\sigma_{x}(t)\rangle$ can be decomposed as 
\be
\langle\sigma_x(t)\rangle=\langle\sigma_{x,0}(t)\rangle+\langle\sigma_{x,\bar{a}}(t)\rangle+\langle\sigma_{x,\bar{b}}(t)\rangle,
\ee
where   $\langle\sigma_{x,0}(t)\rangle$ corresponds to an intial density matrix  without  $\bar{a}$ and $\bar{b}$ while the remaining parts are 
due to the coefficients $\bar{a}$ and $\bar{b}$. 
Similarly to the $z$-component,   for a large cut-off frequency, these two last terms are linked to the even and odd part, with respect to the bias, of 
 $\langle\sigma_{x,0}(t)\rangle$ 
\begin{eqnarray}
\langle\sigma_{x,\bar{a}}(t)\rangle&=& \frac{2\bar{a}}{\Delta_0\sin(\pi\alpha)}\frac{d}{dt}\langle{\sigma^{(+)}_{x,0}(t)}\rangle\nonumber\\
\langle\sigma_{x,\bar{b}}(t)\rangle&=& \frac{2\bar{b}}{\Delta_0\cos(\pi\alpha)(\bar{p}_R-\bar{p}_L)}\frac{d}{dt}\langle{\sigma^{(-)}_{x,0}(t)}\rangle.
\label{Bgenerallinks2}
\end{eqnarray}
We then quote only the series expansion of the term $\langle\sigma_{x,0}(t)\rangle$. We have \cite{uli_book}
\beq
\hspace{-1cm}\langle\sigma_{x}(t)\rangle &=&\langle{\sigma^{(+)}_{x,0}(t)}\rangle+\langle{\sigma^{(-)}_{x,0}(t)}\rangle=
\sum_{n=1}^{\infty}\Delta_0^{2n-1}(-1)^{n-1}2^{-n}\int_0^t dt_{2n-1}\cdots \int_0^{t_2} dt_{1}\cdot\nonumber\\ 
\hspace{-1cm}&&\cdot\sum_{\xi_j=\pm 1}\xi_n\left[{F}_{n,1}^{(-)}C_{n,1}^{(+)}+(\bar p_R-\bar p_L){F}_{n,1}^{(+)}C_{n,1}^{(-)}\right].
\eeq
Comparing now term by term the two series expressions for  $\langle\sigma_{x}(t)\rangle$ and  $\langle\sigma_{z}(t)\rangle$ it is possible to connect them
via an exact integral relation
\be
\hspace{-1cm}\langle \sigma_{x}(t)\rangle = \int_0^t dt' [K^{(+)}_{1,x}(t-t') + K^{(-)}_{1,x}(t-t')\langle \sigma_{z}(t')\rangle ]  +2\bar{a} K^{(+)}_{2,x}(t)+2\bar{b} K^{(-)}_{2,x}(t).
\label{Beq:gme_sigmax}
\ee
Due to the relations (\ref{Bgenerallinks2}), the kernels associated to the $\bar a$ and $\bar b$ are linked to 
$K^{(\pm)}_{1,x} (t)$ as
\begin{eqnarray}
 K_{2,x}^{(+)}(t)&=& \frac{1}{\Delta_0\sin(\pi\alpha)}K_{1,x}^{(+)}(t)\nonumber\\
 K_{2,x}^{(-)}(t)&=& \frac{1}{\Delta_0\cos(\pi\alpha)}K_{1,x}^{(-)}(t).
\label{linksKx}
\end{eqnarray}
The kernels $K^{(\pm)}_{1,x} (t)$ are again obtained by matching the integral relation (\ref{Beq:gme_sigmax}) with the series expansions, the explicit forms are then in terms of  irreducible functionals
\be
\label{Bkx}
\hspace{-2.3cm}K^{(\pm)}_{1,x}(t-t')= {\cal K}^{(\pm)}_{1,x}(t-t')-\sum_{n=2}^{\infty}\Delta_0^{2n-1}\left(\frac{-1}{2}\right)^{n}\int_{t'}^t dt_{2n-1}\cdots \int_{t'}^{t_3} dt_{2}
\sum_{\xi_1\cdots\xi_n=\pm 1} {\xi_n\widetilde{\cal L}}_{n,1}^{(\pm)},
\label{Bkappax}
\ee
with
\beq
\hspace{-2.3cm}{\widetilde{\cal L}}_{n,1}^{(+)}&=&{F}_{n,1}^{(-)}C_{n,1}^{(+)} - \sum_{j=2}^{n}(-1)^j\!\!\sum_{m_1,\cdots,m_j}\!\!{F}_{m_1,1}^{(+)}C_{m_1,1}^{(+)}\cdot{F}_{m_2,1}^{(+)}C_{m_2,1}^{(+)}\cdots {F}_{m_j,1}^{(-)}C_{m_j,1}^{(+)}\cdot
\delta_{m_1+\cdots+m_j,n}\nonumber\\
\hspace{-2.3cm}{\widetilde{\cal L}}_{n,1}^{(-)}&=&{F}_{n,1}^{(+)}C_{n,1}^{(-)}- \sum_{j=2}^{n}(-1)^j\!\!\sum_{m_1,\cdots,m_j}\!\!{F}_{m_1,1}^{(+)}C_{m_1,1}^{(+)}\cdot{F}_{m_2,1}^{(+)}C_{m_2,1}^{(+)}\cdots {F}_{m_j,1}^{(+)}C_{m_j,1}^{(-)}\cdot
\delta_{m_1+\cdots+m_j,n}.\nonumber\\
\hspace{-2.3cm}&&
\eeq
All the symbols were already explained discussing the $z-$component case. The first terms ${\cal K}^{(\pm)}_{1,x}(t-t')$, in (\ref{Bkappax}) are the contributions at order  $\Delta_0$. They reads  
\beq
\label{eq:calk_xa}
{\cal K}^{(+)}_{1,x} (t-t')&=& \Delta_0 e^{-Q'(t-t')}\sin(\pi\alpha) \cos[\epsilon_0(t-t')]\\
\label{eq:calk_xb}
{\cal K}^{(-)}_{1,x} (t-t')&=&  \Delta_0 e^{-Q'(t-t')}\cos(\pi\alpha) \sin[\epsilon_0(t-t')].
\eeq

Before closing this part, we comment on the so-called non-interacting blip approximation (NIBA)~\cite{uli_book, sassetti_1990, carrega_2016}. As already mentioned in the main text, this corresponds to the  lowest order expansion of the kernels in the tunneling amplitude $\Delta_0$.
Therefore the kernels are given by Equations (\ref{eq:calk_za})-(\ref{eq:calk_zb}) and (\ref{eq:calk_xa})-(\ref{eq:calk_xb}) for the $z$- and $x$ components, respectively. For sake of completeness, we report here the closed expressions at order $\Delta_0^2$ for ${\cal K}_{2,z/x}^{(\pm)}$ that can be derived by exploiting the general links explained above.
We have
\beq
\label{eq:calk_z2a}
{\cal K}_{2,z}^{(+)}(t)&=&\Delta_0 e^{-Q'(t)} \cos(\epsilon_0 t)\\
\label{eq:calk_z2b}
{\cal K}^{(-)}_{2,z}(t)&=& \Delta_0  e^{-Q'(t)} \sin(\epsilon_0 t)
\eeq
and
\beq
\label{eq:calk_x2a}
{\cal K}^{(+)}_{2,x} (t)&=& e^{-Q'(t)} \cos(\epsilon_0 t)\\
\label{eq:calk_x2b}
{\cal K}_{2,x}^{(-)} (t)&=&e^{-Q'(t)} \sin(\epsilon_0 t)~.
\eeq
The integro-differential equations posed by~(\ref{eq:gme_sigmaz})-(\ref{eq:gme_sigmax}) can be thus solved numerically using the above NIBA Kernels.
We close by recalling that the NIBA approach is non perturbative in the dissipation strength $\alpha$ and it is a reliable approximation scheme for sufficiently short times~\cite{uli_book, orth_2013}. At long times, it is strictly consistent for temperature higher or of the order of the level splitting $\Omega_{{\rm SB}} =\sqrt{\Delta_0^2 +\epsilon_0^2}$.
\section{Charging dynamics in the NIBA framework}

\label{app:niba}

\begin{figure}[ht]
\centering
\includegraphics[width=0.44\textwidth]{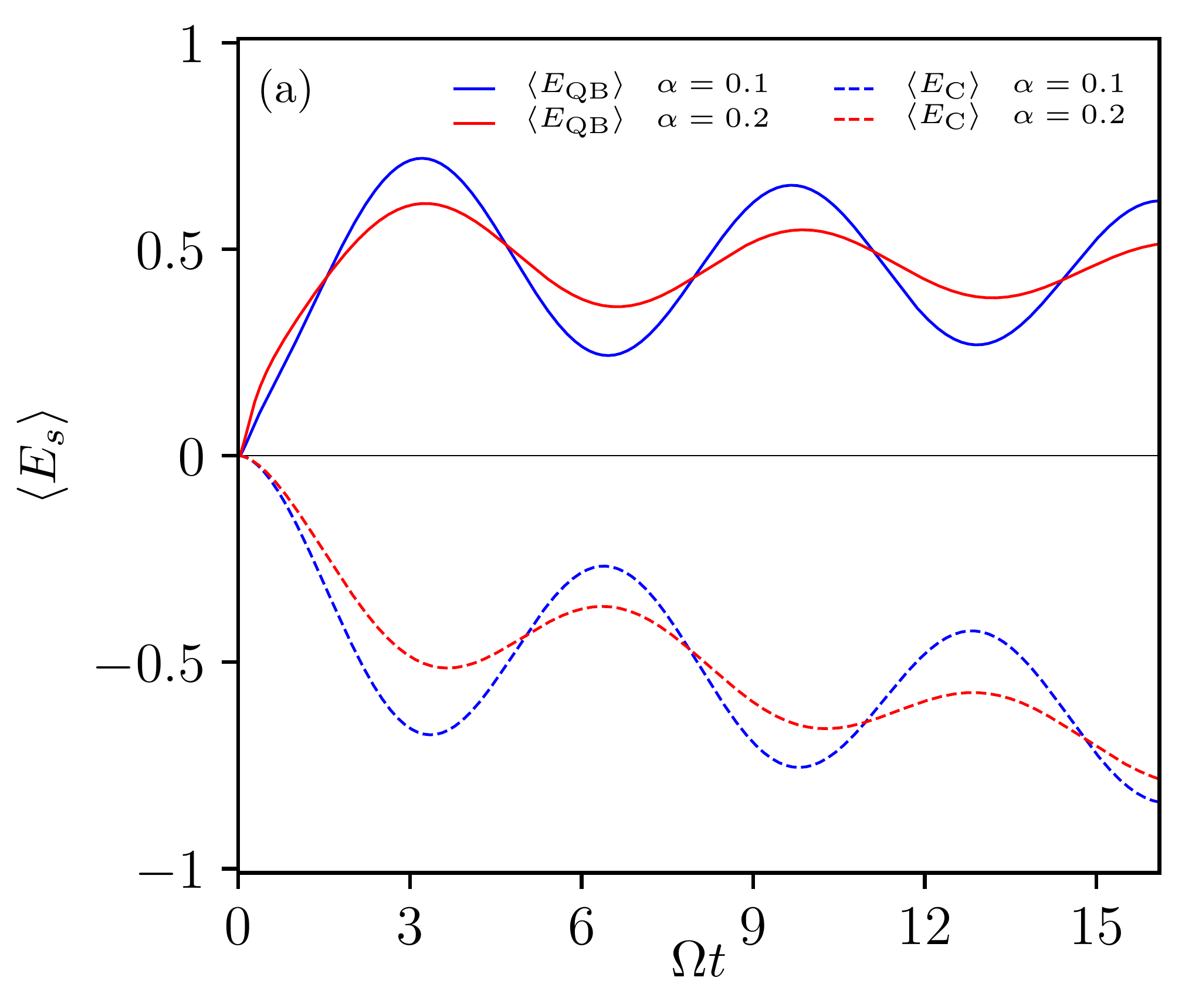}
\includegraphics[width=0.49\textwidth]{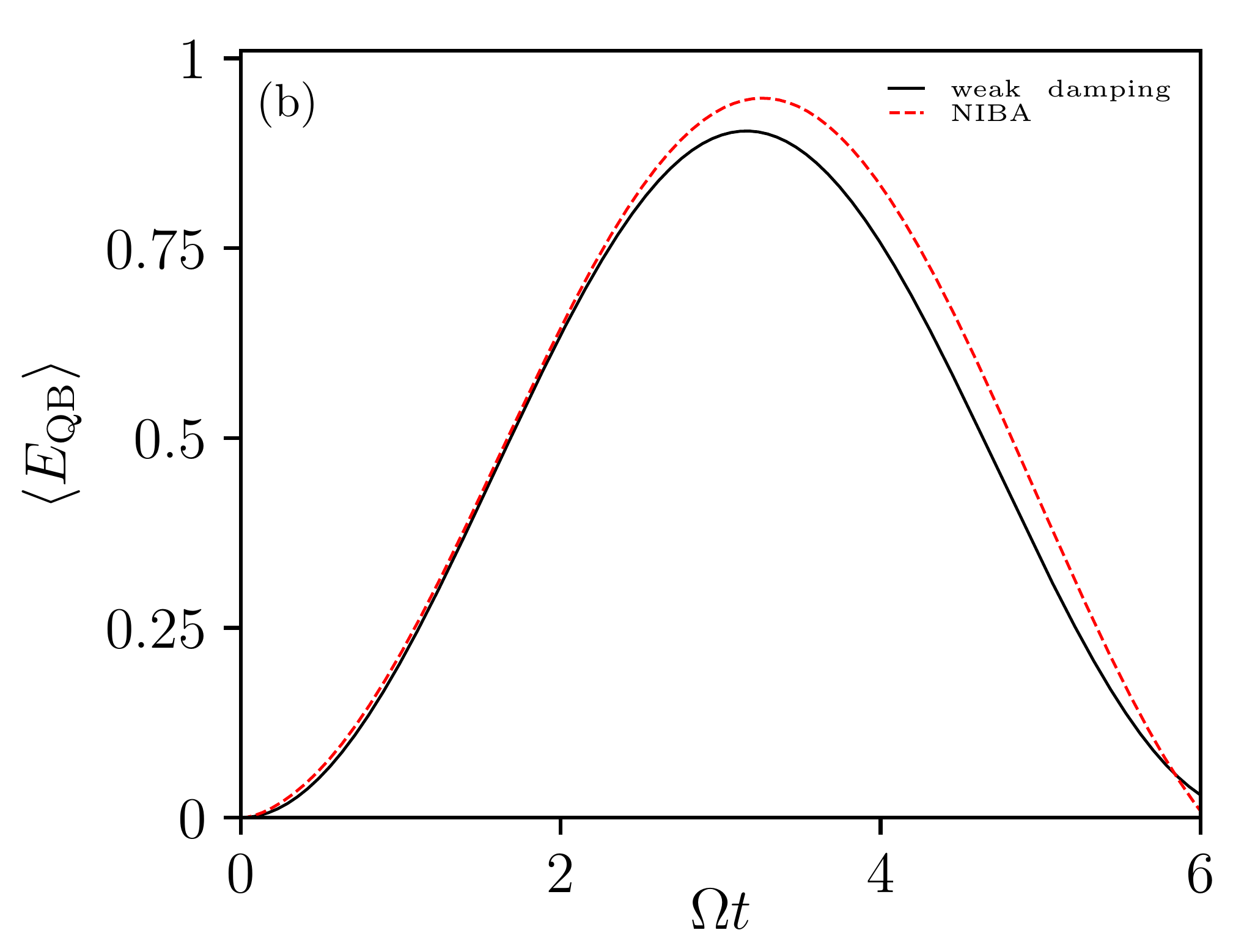}
\caption{Time evolution of average energy stored evaluated in the NIBA framework for $A=3\Delta$ and pure decoherence ($\theta=0$) coupling with the reservoir. Panel (a) shows both energy variations associated to the QB and to the charger for different dissipation strength $\alpha=0.1$ and $\alpha=0.2$. Panel (b) shows comparisons between weak damping expression and the NIBA result at short times for $\alpha=0.01$, demonstrating the validity of NIBA in this short time window. Other parameters as in Figure~\ref{fig1}.}
 \label{fig:niba}
\end{figure}
In this Appendix we present results for the charging dynamics considering higher values of the dissipation strength $\alpha$. These have been obtained within the NIBA framework, by solving the integro-differential equations (\ref{eq:gme_sigmaz})-(\ref{eq:gme_sigmax}), using the closed expressions for the dissipative kernels at order $\Delta_0^2$ reported in Equations~(\ref{eq:calk_za})-(\ref{eq:calk_zb}), (\ref{eq:calk_xa})-(\ref{eq:calk_xb}) and (\ref{eq:calk_z2a})-(\ref{eq:calk_x2b}). The time-evolution of the energy variation of the QB (for both linear dissipative couplings) is then obtained by resorting to the mapping discussed in the main text.
In Figure~\ref{fig:niba} we show results obtained in the NIBA framework for stronger dissipative coupling $\alpha$ with respect to the ones discussed in the main text. As a representative example, we have fixed the driving amplitude at $A=3\Delta$ and we have reported the case of decoherence coupling with the bath in the low temperature regime $\beta\Delta=10$.
As expected, stronger dissipation strengths induce faster relaxation dynamics, resulting in weaker charging performances (see panel (a)). As discussed in the main text, dissipation effects are even stronger in the case of pure dephasing (not shown), whereas in the case shown in Figure~\ref{fig:niba} at short times still a sizeable amount of energy can be stored in the QB for $\alpha=0.1$. Figure~\ref{fig:niba}(b) shows comparison between the NIBA predictions and exact expressions evaluated in the weak damping regime for $\alpha=0.01$, demonstrating that NIBA is still a very good approximation at very short times, even in the low temperature regime considered.


\section*{References}
\begin{thebibliography}{}
%
\bibitem{Esposito09} Esposito M, Harbola U and Mukamel S, 2009 \emph{Rev. Mod. Phys.} \textbf{81} 1665
%
\bibitem{Levy12}
Levy A and Kosloff R 2012  \emph{Phys Rev Lett} {\bf 108} 070604
%
\bibitem{Pekola15} Pekola J 2015 {\it Nat Phys} {\bf 11} 118
%
\bibitem{Vinjanampathy16} Vinjanampathy S and Anders J 2016 \emph{Contemp Phys} \textbf{57} 545
%
\bibitem{Benenti17}
Benenti G, Casati G, Saito K and Whitney R S 2017 \emph{Phys Rep} {\bf 694} 1
%
\bibitem{Depasquale18} De Pasquale A and Stace T M, 2018 \emph{Thermodynamics
in the Quantum Regime}, edited by Binder F, Correa L A, Gogolin C, Anders J, and Adesso G (Springer, Berlin)
%
\bibitem{Bera19} Bera M N, Riera A, Lewenstein M, Khanian Z B and Winter A 2019 \emph{Quantum} \textbf{3} 121
%
\bibitem{Carrega19}
Carrega M, Sassetti M and Weiss U 2019 \emph{Phys Rev} A \textbf{99} 062111
%
\bibitem{Campaioli18} Campaioli F, Pollock F A and Vinjanampathy S, 2018 \emph{Thermodynamics
in the Quantum Regime}, edited by Binder F, Correa L A, Gogolin C, Anders J, and Adesso G (Springer, Berlin)
%
\bibitem{Alicki13} Alicki R and Fannes M 2013 \emph{Phys Rev} E \textbf{87} 042123
%
\bibitem{Hovhannisyan13} Hovhannisyan K V, Perarnau-Llobet M, Huber M and Acin A 2013 \emph{Phys Rev Lett} \textbf{111} 240401
%
\bibitem{Binder15} Binder F C, Vinjanampathy S, Modi K, and Goold J 2015 \emph{New J Phys} \textbf{17} 075015
%
\bibitem{Campaioli17} Campaioli F, Pollock F A, Binder F C, C\'eleri L, Goold J, Vinjanampathy S and Modi K 2017 \emph{Phys Rev Lett} \textbf{118} 150601
%
\bibitem{Julia-Farre18} Juli\'a-Farr\`e S, Salamon T, Riera A, Bera M N and Lewenstein M {\it arXiv:1811.04005}
%
\bibitem{Zhang19} Zhang Y-Y, Yang T-R, Fu L and Wang X 2019 \emph{Phys Rev} E \textbf{99} 052106
%
\bibitem{Chen19} Chen J, Zhan L, Shao L, Zhang X, Zhang Y-Y and Wang X {\it arXiv:1906.06880}
%
\bibitem{Crescente20}
Crescente A, Carrega M, Sassetti M and Ferraro D 2020, ArXiv:2005.05068, in press on New J. Phys.
%
\bibitem{Le18} Le T P, Levinsen J, Modi K, Parish M, and Pollock F A 2018 \emph{Phys Rev} A \textbf{97} 022106
%
\bibitem{Rossini19} Rossini D, Andolina G M, Rosa D, Carrega M and Polini M, {\it arXiv:1912.07234}
%
\bibitem{Rosa19} Rosa D, Rossini D, Andolina G M, Polini M and Carrega M, {\it arXiv:1912.07247}
%
\bibitem{Andolina19b} Andolina G M, Keck M, Mari A, Giovannetti V and Polini M 2019 \emph{Phys Rev} B \textbf{99} 205437
%
\bibitem{Devoret13} Devoret M H, and Schoelkopf R J 2013 \emph{Science} \textbf{339} 1169
%
\bibitem{Ferraro18} Ferraro D, Campisi M, Andolina G M, Pellegrini V and Polini M 2018 \emph{Phys Rev Lett} \textbf{120} 117702
%
\bibitem{Wiel02} van der Wiel W G, De Franceschi S, Elzerman J M, Fujisawa T, Tarucha S, and Kouwenhoven L P 2002 \emph{Rev Mod Phys} \textbf{75} 1
%
\bibitem{Koch07} Koch J, Yu T M, Gambetta J, Houck A A, Schuster D I, Majer J, Blais A, Devoret M H, Girvin S M and Schoelkopf R J 2007 \emph{Phys Rev} A \textbf{76} 042319
%
\bibitem{Singha11} Singha A, Gibertini M, Karmakar B, Yuan S, Polini M, Vignale G, Katsnelson M I, Pinczuk A, Pfeiffer L N, West K W and Pellegrini V 2011 
\emph{Science} \textbf{332} 1176
%
\bibitem{Andolina19} Andolina G M, Keck M, Mari A, Campisi M, Giovannetti V and Polini M 2019 \emph{Phys Rev Lett} \textbf{122} 047702
%
\bibitem{Ferraro19} Ferraro D, Andolina G M, Campisi M, Pellegrini V and Polini M 2019 \emph{Phys Rev} B \textbf{100} 075433
%
\bibitem{Andolina18} Andolina G M, Farina D, Mari A, Pellegrini V, Giovannetti V and Polini M 2018 \emph{Phys. Rev.} B \textbf{98} 205423
%
\bibitem{Haroche_Book} Haroche S and Raimond J-M 2006 \emph{Exploring the quantum. Atoms, Cavities and Photons} (Oxford University Press)
%
\bibitem{uli_book} Weiss U 2012 {\it Quantum dissipative systems}, 4th edition,  (World Scientific, Singapore)
%
\bibitem{Ou17} Ou C, Chamberlin R V and Abe S 2017 \emph{Physica} A \textbf{466} 450
%
\bibitem{Farina19} Farina D, Andolina G M, Mari A, Polini M and Giovannetti V 2019 \emph{Phys. Rev.} B \textbf{99}, 035421
%
\bibitem{Zakavati20} Zakavati S, Tabesh F T and Salimi S, \emph{arXiv:2003.09814}
%
\bibitem{Kamin20} Kamin F H, Tabesh F T, Salimi S, Kheirandish F and Santos A C \emph{arXiv:1910.07751}
%
\bibitem{Barra19} Barra F 2019 \emph{Phys. Rev. Lett.} \textbf{122} 210601 
%
\bibitem{Santos19} Santos A C, \c Cakmak B, Campbell S and Zinner N T 2019 \emph{Phys. Rev.} E \textbf{100} 032107 
%
\bibitem{Gherardini20} Gherardini S, Campaioli F, Caruso F and Binder F C 2020 \emph{Phys. Rev. Research} \textbf{2} 013095
%
\bibitem{Quach20} Quach J Q and Munro W J, \emph{arXiv:2002.10044}
%
\bibitem{schnirman_2002} Schnirman A, Makhlin Y and Schoen G 2002 \emph{Physica Scripta} {\bf{T102}}, 147
%
\bibitem{vion_2002}
Vion D, Aassime A, Cottet A, Joyez P, Pothier H, Urbina C, Esteve D and Devoret M H 2002 \emph{Science} \textbf{296} 886 
%
\bibitem{brandes_2002}
Brandes T and Vorrath T 2002 \emph{Recent Progress in Many-Body Theories} \textbf{471}
%
\bibitem{caldeira_1983} Caldeira A~O and Leggett A~J 1983 {\it Physica A} {\bf 121} 587
%
\bibitem{leggett_1987} Leggett A~J, Chakravarty S, Dorsey A, Fisher M~P, Garg A and Zwerger W 1987 {\it Rev. Mod. Phys.} {\bf 59} 1
%
\bibitem{ingold_2002}
Ingold G~L 2002 Path integrals and their application to dissipative quantum systems {\it Coherent Evolution in Noisy Environments} (Springer) pp 1--53
%
\bibitem{gramich_2011} Gramich V, Solinas P, Mottonen M and Ankerhold J 2011 \emph{Phys. Rev.} A {\bf{84}}, 052103 
%
\bibitem{gramich_2014} Gramich V, Gasparinetti S, Solinas P and Ankerhold J 2014 \emph{Phys. Rev. Lett.} {\bf{113}} 027001 
%
\bibitem{makhlin_2004}
Makhlin Y and Schnirman A 2004, \emph{Phys. Rev. Lett.} {\bf{92}} 178301
%
\bibitem{sassetti_1990} Sassetti M and Weiss U 1990 {\it Phys. Rev. Lett.} {\bf 65} 2262
%
\bibitem{grifoni_1996} Grifoni M, Sassetti M and Weiss U 1996 {\it Phys. Rev.} E {\bf 53} R2033
%
\bibitem{henriet_2014}
Henriet L, Ristivojevic Z, Orth P~P and Le Hur K~L 2014 {\it Phys. Rev.} A {\bf 90} 023820
%
\bibitem{bulla_2003}
Bulla R, Tong N~H and Vojta M 2003 {\it Phys. Rev. Lett.} {\bf 91} 170601
%
\bibitem{orth_2010}
Orth P~P, Imambekov A and Le~Hur K 2010 {\it Phys. Rev.} A {\bf 82} 032118
%
\bibitem{thorwart_2015} Javanbakht S, Nalbach P and Thorwart M 2015 \emph{Phys Rev A} 91, 052103
%
\bibitem{palm_2018} Palm T and Nalbach P 2018, \emph{J. Chem. Phys.} {\bf{149}}, 214103
%
\bibitem{stockburger_2004} Stockburger Y T 2004 \emph{Chem Phys} \textbf{296} 159
%
\bibitem{stockburger_2016} Stockburger Y T 2016 \emph{Euro Phys Lett} \textbf{115} 40010
%
\bibitem{grifoni_1999} Grifoni M, Paladino E and Weiss U 1999 {\it Eur. Phys. J.} B {\bf 10} 719
%
\bibitem{grifoni_1997} Grifoni M, Winterstetter M and Weiss U 1997 {\it Phys. Rev.} E {\bf 56} 334
%
\bibitem{carrega_2015} Carrega M, Solinas P, Braggio A, Sassetti M and Weiss U 2015 \emph{New J. Phys.} {\bf{17}},  045030
%
\bibitem{carrega_2016} Carrega M, Solinas P, Sassetti M and Weiss U 2016, \emph{Phys. Rev. Lett.} {\bf{116}}, 240403 
%
\bibitem{orth_2013} Orth P P, Imambekov and Le Hur K 2013 \emph{Phys Rev B} \textbf{87} 014305
%
\bibitem{hartmann_2000}
Hartmann L, Goychuk I, Grifoni M and Haanggi P 2000 \emph{Phys Rev E} \textbf{61}(R) 4687
%
\bibitem{makhlin_2000}
Makhlin Y, Schoen G and Schnirman A \emph{Rev. Mod. Phys.} {\bf{73}}, 357 (2001)
%
\bibitem{shen_2018}
Shen H Z, Xu S, Yi S and Yi X X 2018 \emph{Phys Rev} A {\bf{98}}, 062106
%
\bibitem{feynman_1963} Feynman R~P and Vernon~Jr F 1963 {\it Ann. Phys.} {\bf 24} 118
%

\end{thebibliography}
\end{document}